\documentclass[longauth]{aa} 

\usepackage{graphicx}
\usepackage{array}
\usepackage{txfonts}
\usepackage[colorlinks=true,linkcolor=blue,citecolor=blue, urlcolor=blue]{hyperref}
\usepackage{longtable}

\begin{document}

   \title{The Pristine survey}
   \subtitle{XXXI. A stream offset from the Stream: very metal-poor stellar debris trailing 100$^{\circ}$ from the LMC-SMC system \thanks{Based on observations gathered with the 8.1m Gemini South Telescope located at the Gemini Observatory, Chile.}}
   \titlerunning{A stream offset from the Stream}

   \author{Akshara Viswanathan
          \inst{\ref{uvic},\ref{paris}}\fnmsep\thanks{Corresponding author; \email{astroakshara97@gmail.com}}
          \and
          Kim~A.~Venn\inst{\ref{uvic}}
          \and
          Julio~F.~Navarro\inst{\ref{uvic},\ref{korea}}
          \and
          Daniel~A.~Boyea\inst{\ref{uvic},\ref{kapteyn}}
          \and
          Federico~Sestito\inst{\ref{hert}}
          \and
          Camila~Navarrete\inst{\ref{chile}}
          \and
          Hanneke~C.~Woudenberg\inst{\ref{kapteyn}}
          \and
          Tadafumi~Matsuno\inst{\ref{heidelberg}}
          \and
          Vinicius~M.~Placco\inst{\ref{noirlab}}
          \and
          Madeleine~McKenzie\inst{\ref{carnegie}}
          \and
          Triana~Cornejo\inst{\ref{uvic}}
          \and
          Anya~Dovgal\inst{\ref{uvic}}
          \and
          Anke~Ardern-Arentsen\inst{\ref{cambridge}}
          \and
          Manuel~Bayer\inst{\ref{kapteyn}}
          \and 
          Amanda~Byström\inst{\ref{edin}}
          \and
          Vanessa~Hill\inst{\ref{nice}}
          \and
          Pascale~Jablonka\inst{\ref{epfl},\ref{paris}}
          \and 
          Jaclyn~Jensen\inst{\ref{uvic},\ref{dartmouth}}
          \and 
          Georges~Kordopatis\inst{\ref{nice}}
          \and
          Nicolas~F.~Martin\inst{\ref{strasbourg},\ref{mpia}}
          \and
          Rubén~Sánchez-Janssen\inst{\ref{ing},\ref{iac},\ref{edin}}
          \and
          Simon~E.~T.~Smith\inst{\ref{uvic}}
          \and
          Else~Starkenburg\inst{\ref{kapteyn}}
          \and
          Daria~Zaremba\inst{\ref{uvic},\ref{mit}}
          \and
          Giuseppina~Battaglia\inst{\ref{iac},\ref{lalaguna}}
          \and
          Salvatore~Taibi\inst{\ref{epfl}}
          }

   \institute{Dept. of Physics and Astronomy, University of Victoria, P.O. Box 3055, STN CSC, Victoria BC V8W 3P6, Canada\label{uvic}
         \and
         LIRA, Observatoire de Paris, Université PSL, Sorbonne Université, Université Paris Cité, CY Cergy Paris Université, CNRS, F-92190 Meudon, France\label{paris}
         \and
         Crimson Distinguished Professor, KU-KIST Green School, Korea University, Seoul, South Korea\label{korea}
         \and
         Kapteyn Astronomical Institute, University of Groningen, Landleven 12, 9747 AD Groningen, The Netherlands\label{kapteyn}
         \and
         Centre for Astrophysics Research, University of Hertfordshire, Hatfield, AL10 9AB, UK\label{hert}
         \and
         Departamento de F\'isica, Universidad T\'ecnica Federico Santa Mar\'ia, Vicu\~{n}a Mackenna 3939, San Joaqu\'in, Santiago de Chile, Chile\label{chile}
         \and
         Astronomisches Rechen-Institut, Zentrum für Astronomie der Universität Heidelberg, Mönchhofstraße 12-14, 69120, Heidelberg, Germany\label{heidelberg}
         \and
         NSF NOIRLab, Tucson, AZ 85719, USA\label{noirlab}
         \and
         Observatories of the Carnegie Institution for Science, 813 Santa Barbara St., Pasadena, CA 91101, USA\label{carnegie}
         \and
         Institute of Astronomy, University of Cambridge, Madingley Rd, Cambridge CB3 0HA, UK\label{cambridge}
         \and
         Institute for Astronomy, University of Edinburgh, Royal Observatory, Blackford Hill, Edinburgh EH9 3HJ, UK\label{edin}
         \and
         Université Côte d’Azur, Observatoire de la Côte d’Azur, CNRS, Laboratoire Lagrange, Nice, France\label{nice}
         \and
         Institute of Physics, Laboratory of Astrophysics, Ecole Polytechnique Fédérale de Lausanne (EPFL), 1290 Sauverny, Switzerland\label{epfl}
         \and
         Department of Physics \& Astronomy, Dartmouth College, 6127 Wilder Laboratory, Hanover, NH 03755, USA\label{dartmouth}
         \and
         Université de Strasbourg, CNRS, Observatoire astronomique de Strasbourg, UMR 7550, F-67000 Strasbourg, France\label{strasbourg}
         \and
         Max-Planck-Institut für Astronomie, Königstuhl 17, D-69117 Heidelberg, Germany\label{mpia}
         \and
         Isaac Newton Group of Telescopes, Apartado 321, E-38700 Santa Cruz de la Palma, Tenerife, Spain\label{ing}
         \and
         Instituto de Astrofísica de Canarias, calle Vía Láctea s/n, E-38205 La Laguna, Tenerife, Spain\label{iac}
         \and
         Department of Physics and MIT Kavli Institute for Astrophysics and Space Research, Massachusetts Institute of Technology, 77 Massachusetts Avenue, Cambridge, MA 02139, USA\label{mit}
         \and
         Universidad de La Laguna, Dpto. Astrofísica, E-38206, La Laguna, Tenerife, Spain\label{lalaguna}
             }

   \date{Received \today; accepted xxx}

 
  \abstract
    {The stellar counterpart to the $200^\circ$-long H\,{\sc i} gas Stream emanating from the Clouds provides a direct probe of the stellar populations tidally stripped from them. Its spatial, kinematic, and chemical properties offer important constraints on the interaction history of the Clouds and their structure prior to infall into the Milky Way.}
    {We investigate the chemodynamical properties and possible origin of very metal-poor (VMP; [Fe/H] $<-2$) stars in the stellar counterpart to the gas Stream, targeting distant red giants selected from the Pristine--Gaia synthetic photometric metallicity catalogue.}
    {We obtained high-resolution Gemini/GHOST spectroscopy of 33 candidate members of the stellar stream of the Clouds, deriving radial velocities and spectroscopic metallicities. We combine these observations with previously identified red giant members from the literature and investigate their dynamical association with the Clouds using orbital integrations in a time-dependent Milky Way--LMC potential. We also compare their distribution with forward-integrated collisionless test particles initially distributed in a diffuse stellar halo around the LMC.}
    {We identify 12 high-confidence members using their energy and angular momentum: one intermediate metal-poor star ([Fe/H] $\sim-1.6$), ten VMP stars, and one extremely metal-poor (EMP) star ([Fe/H] $<-3$). The newly identified VMP stars trace the trailing stellar Stream over more than $100^\circ$ on the sky, following the kinematics of the Clouds while being spatially offset by $\sim20^\circ$ from both the H\,{\sc i} gas Stream and previously identified, more metal-rich stellar debris. This suggests a metallicity-dependent spatial distribution within the stellar Stream. Together with the literature sample, these discoveries approximately double the known red giant member sample and extend its metallicity distribution from [Fe/H] $\sim-0.5$ to $-4.3$. Orbital integrations support an association of the VMP stars with the Clouds, while the test-particle simulations reproduce the broad phase-space distribution of the observed stellar debris.}
    {The presence of VMP and EMP stars in the Stream demonstrates that metal-poor populations associated with the Clouds contribute to their tidal debris. These stars may originate in the extended stellar halos of the LMC or the SMC, or in disrupted satellites associated with the Clouds. Their spatial and kinematic properties provide important constraints for future simulations of the Milky Way--LMC--SMC interaction. }


   \maketitle
%
\section{Introduction}

The Clouds are one of the most remarkable laboratories for studying galaxy interactions in the nearby Universe. 
The large and small of these dwarf galaxies (LMC and SMC), the Bridge, the Leading Arm, and the Stream together provide a unique opportunity to investigate tidal interactions, gas accretion, circumgalactic medium (CGM) physics, and the assembly history of the Milky Way. 
Among these substructures of the Clouds, the gas M. Stream (MS, or the Stream) is the most prominent. 
It is an enormous gaseous feature extending over more than 200$^{\circ}$ across the southern sky and contains both neutral and ionized gas distributed far beyond the present-day extent of the Clouds \citep[see][for a detailed review]{2016donghia, 2026besla}.

Observations over the last five decades have revealed a complex morphology for the MS. 
Early H\,{\sc i} surveys identified both a trailing stream extending behind the Clouds and a Leading Arm located ahead of their orbit  \citep{1965dieter,1972wannier, 1974mathewson, 1998putman}. 
The trailing stream itself appears bifurcated into multiple filaments that differ in kinematics and chemical composition, suggesting contributions from both the LMC and SMC \citep{2003putman,2008nidever,2013fox,2013richter}. 
Despite extensive observational and theoretical efforts, the origin of the MS remains an open question. 
Two broad mechanisms have been proposed for the formation of the gas Stream: (i) tidal stripping during past interactions between the Clouds, which can produce both gaseous and stellar debris, and (ii) ram-pressure stripping as the Clouds move through the Milky Way's gaseous halo, which removes gas without stripping pre-existing stars. 
Both processes likely play a role, but their relative contributions remain uncertain \citep{2012besla,2013besla,2020lucchini,2022krishnarao}.


Several key questions concerning the MS remain unresolved. 
The true origin of the Leading Arm is still debated, with some studies favouring a tidal origin involving material stripped from the Clouds, while others suggest that ram-pressure stripping of gas from dwarf galaxies, unrelated to the Clouds, and moving ahead of the Clouds may also contribute \citep{2014yang,2015hammer,2015tepper,2019tepper,2022bustard}. 
The distance to the gaseous stream remains poorly constrained because direct distance measurements to diffuse gas are not possible \citep{2017barger,2020antwi}. 
The ionization mechanism responsible for the large reservoir of ionized gas associated with the stream is likewise uncertain, with possible contributions from Galactic radiation fields, shock heating, and past activity from the Galactic nucleus \citep{2007bland,2013bland,2017barger}. 
These uncertainties directly affect estimates of the total mass of the Clouds and the amount of material being accreted onto the Milky Way.

A particularly promising avenue for addressing these questions is the stellar counterpart of the gas Stream, the M. Stellar Stream (MSS\footnote{Throughout this paper, we use ``the Clouds'' collectively, and LMC and SMC separately \citep{2026besla}. We use the abbreviations MS and MSS for the gas and stellar Streams, respectively. We acknowledge that the Clouds have long been known to Indigenous people of the Southern Hemisphere and feature in their oral traditions \citep{2020dennefeld}; they were neither discovered nor named by Magellan. The conventional name nevertheless commemorates him, overlooking these Indigenous histories and the violence he inflicted on Indigenous peoples in South America \citep{2023delosreyes}.}). 
Most tidal models predict that stars should have been stripped from the Clouds alongside the gas \citep{1995lin,1996gardiner,2011diaz,2012diaz,2012besla,2013besla}. 
Such a stellar stream would provide direct distance measurements, precise kinematics, and detailed chemical abundance information that are difficult or impossible to obtain from the gas alone. 
The existence and properties of the MSS therefore place strong constraints on the interaction history of the Clouds, the formation mechanism of the stream, and the gravitational potential of both the Milky Way and the Clouds \citep{2019garavito,2021garavito,2021vasiliev,2021Petersen,2024vasiliev}.

Despite decades of searching, identifying an unambiguous stellar counterpart to the MS has proven difficult \citep{1982recillas,1983brueck,1997zaritsky}. 
Various studies have reported candidate stellar structures aligned with the MS, including blue horizontal branch (BHB) stars, red giant branch (RGB) stars, and RR Lyrae stars in the Galactic North and South \citep{2016belokurov,2020zaritsky,2022petersen}. 
However, establishing a definitive association requires full phase-space information and chemical abundances because the outer halo contains numerous overlapping substructures. 
The recent discovery of stars with kinematics and chemistry consistent with a Clouds origin has provided some of the strongest evidence to date for the existence of an MSS \citep{2023chandra}. 
Nevertheless, the currently known sample remains sparse, and the global extent, metallicity distribution, and dynamical properties of the stellar stream counterpart are still poorly characterized. 
The angular offset ($\sim$20$^{\circ}$) between the gaseous stream and the metal-poor (MP; $[\mathrm{Fe/H}]<-1.5$) subset of the stellar stream identified by \citet{2023chandra} is also noteworthy.

The MSS is also interesting in the broader context of Milky Way halo studies. 
The outer halo contains the debris of past accretion events, dynamical perturbations induced by the LMC, and structures associated with the LMC wake \citep{2019belokurov,2021conroy,2025cavieres}. 
Distinguishing between stars directly stripped from the Clouds and stars affected by the Milky Way's dynamical response to LMC's infall is essential for understanding both populations. 
Furthermore, metal-poor members of the MSS likely represent some of the oldest stellar populations in the outskirts of the Clouds, providing a unique window into their early chemical evolution and assembly history \citep{2025placco,2026viswanathan}.

The recent identification of a population of very metal-poor stars tentatively associated with the MSS by \citet{2026viswanathan} motivates a spectroscopic investigation of the metal-poor tidal debris of the LMC--SMC. 
While the candidate members identified in that work exhibit positions, and proper motions suggestive of a Clouds origin, spectroscopic radial velocities, and distances are required to establish their dynamical association with the Clouds. 
In this paper, we present Gemini South/GHOST \citep{2024connachie, 2024kalari} spectroscopic observations of targets selected by \citet{2026viswanathan}. 
By combining these observations with Gaia astrometry and literature samples of candidate MSS red giant stars \citep{2023chandra,2025placco,2025limberg,2026ji}, we investigate the dynamical properties of stars potentially associated with the Clouds across a broad range of metallicities and distances. 
This allows us to construct the most homogeneous dynamical analysis of the MSS to date, going down to the lowest metallicities.

The paper is organized as follows. 
In Section \ref{methods}, we describe the target selection, Gemini/GHOST observations, data reduction, and the derivation of radial velocities and metallicities. 
In Section \ref{results}, we present the kinematic and orbital properties of the observed stars, identify the subset most likely associated with the Clouds, and combine our sample with previously reported MSS candidates from the literature to construct a unified view of the MSS and compare the resulting sample with a test-particle model of a diffuse stellar halo around the LMC. 
Finally, in Section \ref{conclusions}, we discuss the implications of our results on the origin and extent of the MSS, summarize our main conclusions, and outline directions for future work.




\section{Data and Methods}\label{methods}

\subsection{Target selection}

Our MSS candidates were selected following \citet{2026viswanathan}, using Gaia astrometry, photometry, and distance constraints ($d>40$ kpc). 
Candidates were required to lie within a certain Galactic region ($b(l)=-60^\circ+0.1l+0.007l^2\pm20^\circ$) where the outer halo overdensity was discovered by \citet{2024viswanathan}, with proper-motions aligned towards the Clouds. 
All targets serendipitously have photometric [Fe/H] $\lesssim-2.0$, due to the distance--metallicity selection effect which favours the most metal-poor stars, at larger distances as discussed in Section~3.6.2 of \citet{2026viswanathan}. 


\subsection{Observations}

We observed MSS candidates with the Gemini High-resolution Optical SpecTrograph (GHOST) on Gemini South through programs GS-2025B-Q-310 (PI Viswanathan) and GS-2023B-FT-104, GS-2024A-Q-221, and GS-2024A-Q-319 (PI Sestito). GHOST is a fibre-fed, dual-channel high-resolution spectrograph providing simultaneous blue- and red-arm spectra \citep{2024kalari,2024connachie}. 
Of the 41 candidates in our target list, one with $\delta>30^\circ$ was inaccessible from Gemini South, and 33 out of the remaining 40 were observed. 
We used the standard-resolution mode ($R\sim56\,000$) in single-IFU configuration, with exposure times tailored to the targets ($G\sim15.5$--17.3) to achieve at least S/N$>10$ in the Calcium \,{\sc ii} Triplet (CaT, $\sim$8500\AA) region. 
The data were obtained with $2\times8$ (spectral$\times$spatial) binning. 
Observation details are provided in Appendix Table~\ref{appendixtable2}.

\subsection{Data reduction and radial velocities}

The GHOST spectra were reduced using the Gemini DRAGONS pipeline \citep{2023labrie,2024placco}, which performs standard CCD processing, wavelength calibration, and an initial barycentric correction. 
The resulting one-dimensional spectra from the blue and red arms were further processed using \texttt{GHOULS\_FIXER}\footnote{\href{https://github.com/anyadovgal/GHOULS_FIXER_v3}{github.com/anyadovgal/GHOULS\_FIXER\_v3}} as follows. 
Individual exposures were continuum-normalised using an automated smoothing and iterative sigma-clipping procedure, with the blue arm treated in two wavelength segments (3700--4405 and 4395--5305\,\AA) and the red arm as a single segment (5295--9500\,\AA). 
Radial velocities (RV) were measured for each exposure by cross-correlating the spectra with a GHOST observation of the metal-poor standard star HD~122563, using both broad wavelength regions and masks around prominent diagnostic lines (Ca\,H\&K, Balmer, Mg\,b, Na\,D, and CaT). 
Formal RV uncertainties were estimated from the cross-correlation peak variance, and velocities were corrected to the Solar System barycentre using the telescope coordinates at the time of observation. 
The individual exposures were then shifted to the rest frame and co-added using IRAF \texttt{scombine} routine \citep{1986iraf,1993iraf,2025iraf} to produce the final spectra for the chemodynamical analysis.

\subsection{Distances}

Reliable distances are required for the CaT metallicity calibration which depends on stellar luminosity, and for dynamical analysis. 
At the large distances of our MSS candidates ($d\gtrsim30$--60 kpc), Gaia parallaxes are generally non-informative, making direct parallax inversion inappropriate. 
The isochrone-based distances in \citet{2026viswanathan} rely on spectro-photometric parameters inferred from Gaia XP spectra \citep{2023andrae}, which can be affected by systematics at low metallicities and faint magnitudes. 
We therefore adopt a Bayesian framework following \citet{2019sestito}, combining Gaia astrometry and broadband photometry with a halo-informed spatial prior \citep{2021bailerjones}. 
The parallax contributes weakly to the posterior, which is primarily constrained by photometry. 
We further use MESA/MIST isochrones for old populations (ages 11--12 Gyr), and the Pristine-Gaia photometric metallicity of the star to break the distance--luminosity degeneracy. 
The resulting posteriors favour giant-branch solutions for all targets, consistent with our RGB target selection.
Although the typical fractional distance uncertainties are $\sim$8--12\%, we adopt a conservative 20\% uncertainty for all stars to account for systematic effects in photometric distances to very metal-poor outer-halo stars \citep[see for e.g.][]{2025placco}. 

%

\subsection{Metallicities}

We derive spectroscopic metallicities by modelling the three near-infrared CaT lines (8498.04, 8542.09, and 8662.14\,\AA) in the normalised, rest-frame GHOST spectra, using a modified version of the MCMC pipeline developed by \citet{2024viswanathan}. 
Each line is fitted with a Voigt profile, and the MCMC algorithm samples the posterior distributions of the line depths and widths simultaneously. 
We use CaT-based spectroscopic metallicities, because it provides a robust diagnostic for RGB stars in the very metal-poor regime \citep[e.g.,][]{2022li,2022longeard,2023sestito,2025viswanathan,2026geha,2026bayer}.

Converting the CaT equivalent widths to metallicities requires extinction-corrected photometry and stellar distances. 
We use the iterative extinction-correction of \citet{2024martin}, transforming the dereddened Gaia photometry into Johnson--Cousins $I$ magnitudes following \citet[Table C.2]{2021riello}. 
We then use the empirical calibration of \citet{2013carrera}, which relates CaT equivalent widths to [Fe/H] as a function of stellar luminosity. 
The resulting metallicities are consistent with those obtained using the \citet{2010starkenburg} calibration with Gaussian-fitted equivalent widths and a correction for the line wings.

\begin{figure}
   \centering
   \includegraphics[width=\linewidth]{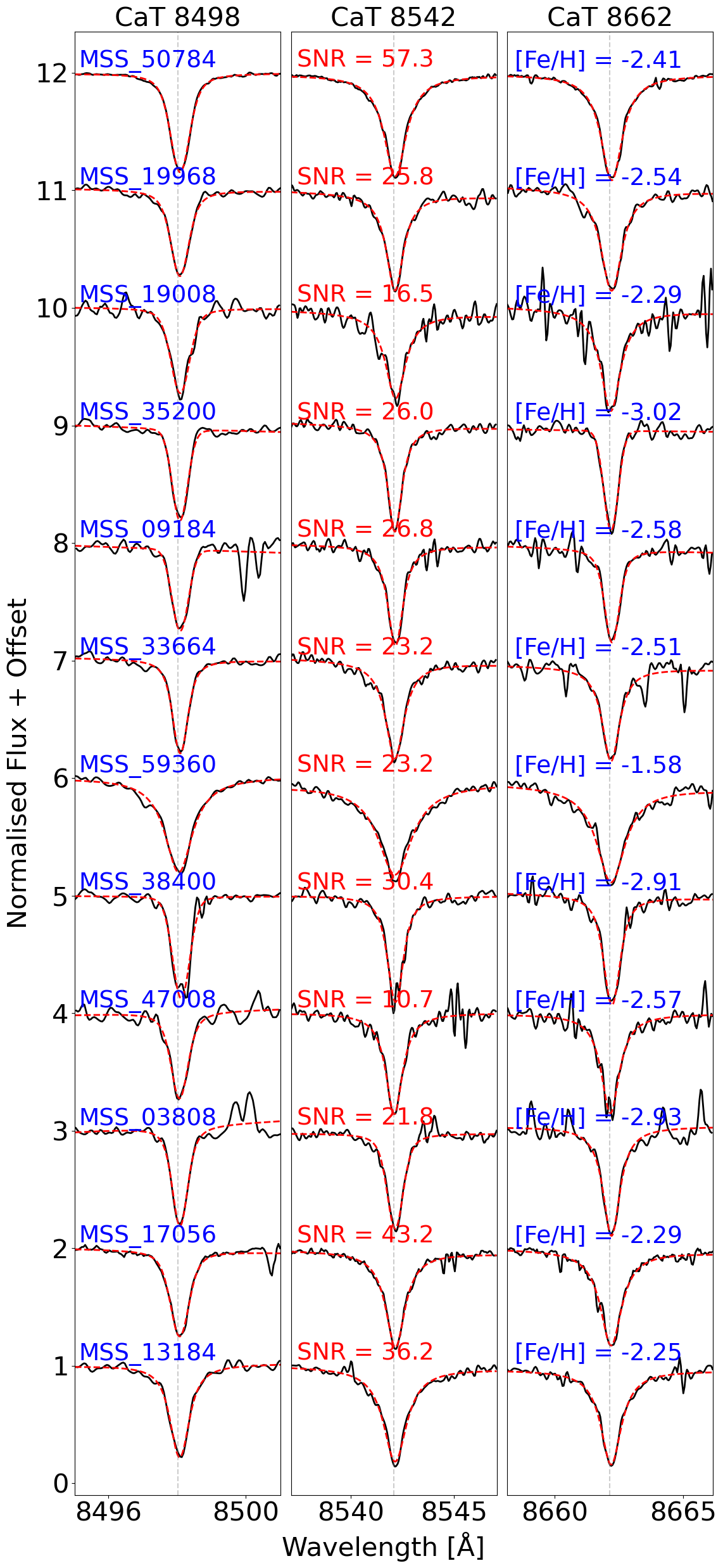}
   \caption{CaT spectral fits for 12 (out of the 33 candidates) MSS members (see section \ref{e-lz-selection} for the selection criteria) observed with GHOST. The three columns show the three CaT features at 8498.04 \AA, 8542.09 \AA, and 8662.14 \AA, respectively. The normalised rest-frame spectra are shown in black and vertically offset for clarity, while the best-fitting Voigt profile models are shown in orange. The target names, signal-to-noise ratios measured in the CaT region, and derived spectroscopic metallicities are indicated in the left, middle, and right panels, respectively. The sample spans a metallicity range from [Fe/H] $\sim$ --1.6 to [Fe/H] $\sim$ --3.1, with 11 out of 12 stars being VMP, demonstrating our CaT Voigt-profile fitting procedure works very well across the very metal-poor regime probed by the MSS candidates.}
              \label{fig:CaT_fits}%
\end{figure}

Metallicity uncertainties are estimated by propagating errors in equivalent widths, photometry, colour transformations, and distances through 1000 Monte Carlo realisations per star. 
We report the mean and standard deviation of the resulting metallicity distributions, approximated as Gaussian. 
This yields CaT-based metallicities for all 33 observed candidates and enables a chemodynamical analysis of the MSS at the VMP end. 
The spectral fits are illustrated in Figure~\ref{fig:CaT_fits}, and a comparison with the photometric metallicities of the Pristine--Gaia catalogue is presented in Appendix~\ref{app:photmet}.

\section{Results and Discussion}\label{results}

To place our Gemini/GHOST sample of MSS candidates in the broader context of stars previously suggested to be associated with the Clouds, we add to our sample all such currently known RGB stars in the literature for which full 6D phase-space information and spectroscopic metallicities are available. 
This comparison sample comprises 15 stars drawn from four recent studies. We include 11 MSS stars from \citet{2023chandra}, selected from their sample of distant halo giants. 
Two of their most metal-poor stars (Gaia DR3 IDs: 6611869850296545024, 6619464348908375808) are excluded from our analysis because their positions, proper motions, and radial velocities overlap with the Orphan-Chenab (OC) stellar stream \citep{2023koposov}. 
The remaining \citet{2023chandra} metal-poor ([Fe/H]<--1.5) stars that overlap with the OC stream exhibit consistently higher fractions of LMC-bound orbital realizations as shown in Table 1 in \citet{2023chandra} and are therefore retained in our analysis. 
We further include the three EMP stars identified by \citet{2025placco} from the spectroscopic follow-up of DECam MAGIC survey \citep{2026chiti} EMP candidates, together with the ultra metal-poor star (UMP, [Fe/H]=--4.3) discovered independently by \citet{2025limberg} and \citet{2026ji}, from follow-up spectroscopy of Gaia XP-selected metal-poor candidates and SDSS-V EMP stars respectively. 
These objects lie very close to the LMC on the sky, and are on the leading side of the LMC.

We do not include the 16 candidate RGB stars identified by \citet{2020zaritsky} near the trailing arm using the H3 Survey \citep{2019conroy}, because of the substantial difference in their distance estimates using two different methods which will complicate our dynamical analysis, and individual spectroscopic metallicities are not reported for these stars.


\subsection{Energy and angular momentum}\label{e-lz-selection}

\begin{figure*}
   \centering
   \includegraphics[width=\linewidth]{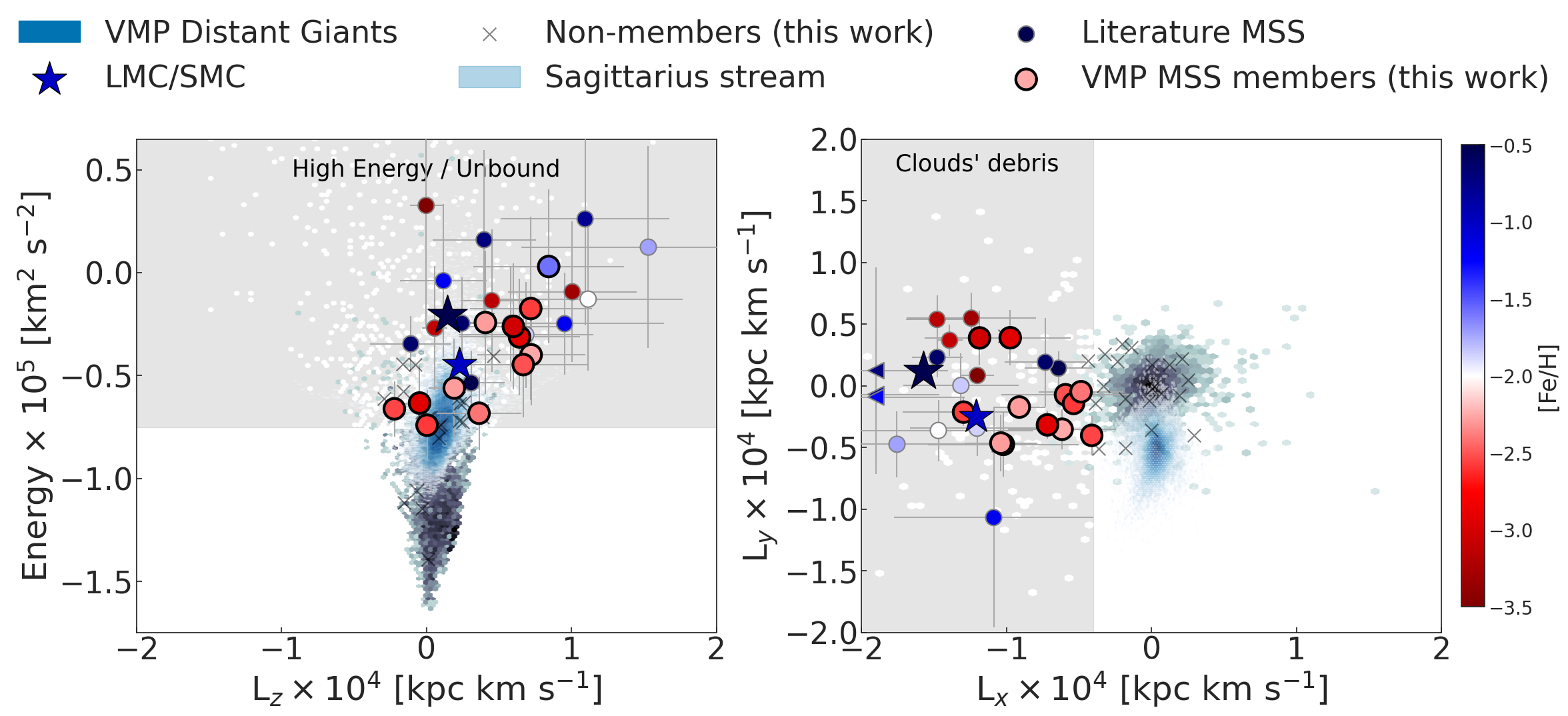}
   \caption{Selection of MSS members using Energy and Angular Momentum. Left: Energy in \texttt{PriceWhelan2022} Milky Way potential versus angular momentum in z-direction. Right: Angular momentum in y-direction versus x-direction. VMP red giant stars in the outer halo from \citet{2026viswanathan} are shown in gray, and the Sagittarius stream red giant stars from \citet{2021vasiliev} are shown in blue to show that these populations have different energy and angular momentum distribution compared to our MSS stars. The Clouds (shown as large star symbols) are orbiting in the y-z orbital plane and so are the MSS members from the literature (shown as small circles) and our VMP MSS members (shown as large circles) colour-coded by their spectroscopic metallicities. These stars are also occupying high-energy orbits compared to the VMP halo population. Non-members from our spectroscopic follow-up are shown as black crosses overlapping with the isotropic VMP halo population. We use a canonical L$_x$<--4$\times$10$^3$ kpc km s$^{-1}$ and E>--0.75$\times$10$^5$ km$^2$ s$^{-2}$ as the MSS member selection.}
   \label{e-lz}%
\end{figure*}

Merger debris from past accretion events that assembled the Milky Way stellar halo remain clustered in integrals-of-motion (IOM) space over long dynamical timescales \citep{2020helmi}. 
In a static or slowly evolving gravitational potential, the total orbital energy ($E$) and components of the angular momentum vector ($\mathbf{L} = [L_x, L_y, L_z]$) serve as robust phase-space diagnostics to isolate tidal structures from the smooth background halo.

To evaluate the orbital properties of our target stars, we perform Galactocentric coordinate transformations and orbit integration using the \texttt{Agama} dynamics framework \citep{2019vasiliev}.
We use a right-handed Galactocentric frame, adopting a solar position of $\mathbf{x}_\odot = (-8.12, 0.00, 0.02) \text{ kpc}$ and a solar velocity of $\mathbf{v}_\odot = (12.9, 245.6, 7.8) \text{ km s}^{-1}$ \citep{2018gravity}.
Specific total energies $E = \frac{1}{2} v^2 + \Phi(\mathbf{r})$ are computed using the Milky Way potential implemented in the \texttt{gala} package \citep{2022pricewhelan}. 
This Galactic potential is represented by the four massive components, including a nucleus, a Dehnen bulge ($M = 5 \times 10^9\,M_\odot$), a three-component Miyamoto--Nagai approximation to an exponential disc ($M_{\mathrm{disk}} = 4.77 \times 10^{10}\,M_\odot$), and a spherical NFW halo ($r_s = 15.6\,\text{kpc}$, $M_{<5.3\,r_s} = 5.54 \times 10^{11}\,M_\odot$).

Because the Clouds orbit the Milky Way in a near-polar plane dominated by $y$--$z$ motion, stars stripped from the Clouds inherit a distinct angular momentum orientation (specifically, strongly negative $L_x$) and higher orbital energies relative to the smooth halo, owing to the LMC's first infall passage. Figure~\ref{e-lz} compares the orbital properties of our targets with the baseline distribution of very metal-poor (VMP) outer-halo ($d>25$ kpc) giants from \citet{2026viswanathan}. The isotropic background VMP halo exhibits a symmetric distribution centred near $L_x \approx 0$ and lower mean orbital energies.

To establish empirical boundaries for membership selection, we use the orbital energy and angular momentum distributions shown in Figure~\ref{e-lz}. We set our selection thresholds at the $\approx 1\sigma$ offset boundary of the background VMP halo distribution:
\begin{equation}
    L_x < -4 \times 10^3\,\text{kpc\,km\,s}^{-1} \quad \text{and} \quad E > -0.75 \times 10^5\,\text{km}^2\,\text{s}^{-2}.
\end{equation}
Stars satisfying both criteria—comprising the 15 literature candidates and 12 out of 33 stars observed in this work are clearly separated from the background in Figure \ref{e-lz}. 
This criteria also selects two more stars (Gaia DR3 ID: 6779255411703094656, 6785860826587863040) from our follow-up that are likely OC stream contaminants and are therefore, not included in our final MSS sample.
The trailing arm of the OC stream overlaps with the MS footprint in both \citet{2023chandra} and \citet{2026viswanathan}, with similar metallicities (mean metallicity of -1.93 and dispersion of about 0.4), and similarly negative $L_x$ angular momentum (median $Lx$ $\sim$ $-3 \times 10^3\,\text{kpc\,km\,s}^{-1})$. 
Therefore, any MSS candidate close to the OC stream in position, proper motions, and radial velocities are removed from the final MSS selection.
The origin of the remaining 21 targets, classified as non-members, are discussed in Appendix~\ref{app:nonmembers}. 

\subsection{The Stellar Stream across the entire metallicity range}\label{members}

\begin{figure*}
   \centering
   \includegraphics[width=\linewidth]{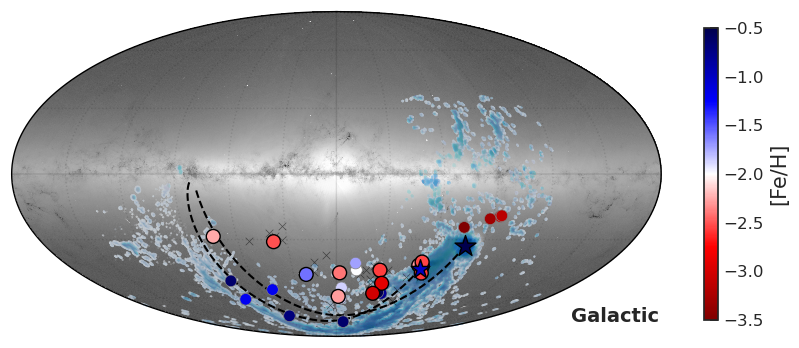}
   \caption{MSS members overplotted on top of the Gaia DR3 stellar density map \citep{2023gaia} and H\,{\sc i} column density map of the gas Stream \citep{2010nidever}. Literature stars from \citet{2023chandra}, \citet{2025placco}, and \citet{2025limberg,2026ji} are shown as small circles, the VMP MSS stars from this work are shown as large circles colour-coded by their spectroscopic metallicities, and the LMC/SMC themselves as large/small stars color-coded by their mean metallicities. Non-members from our spectroscopic follow-up are shown as black crosses. The dashed lines trace the past 1 Gyr orbit of the Clouds \citep{2024vasiliev}}
              \label{mss-sky}%
    \end{figure*}

    \begin{figure*}
   \centering
   \includegraphics[width=\linewidth]{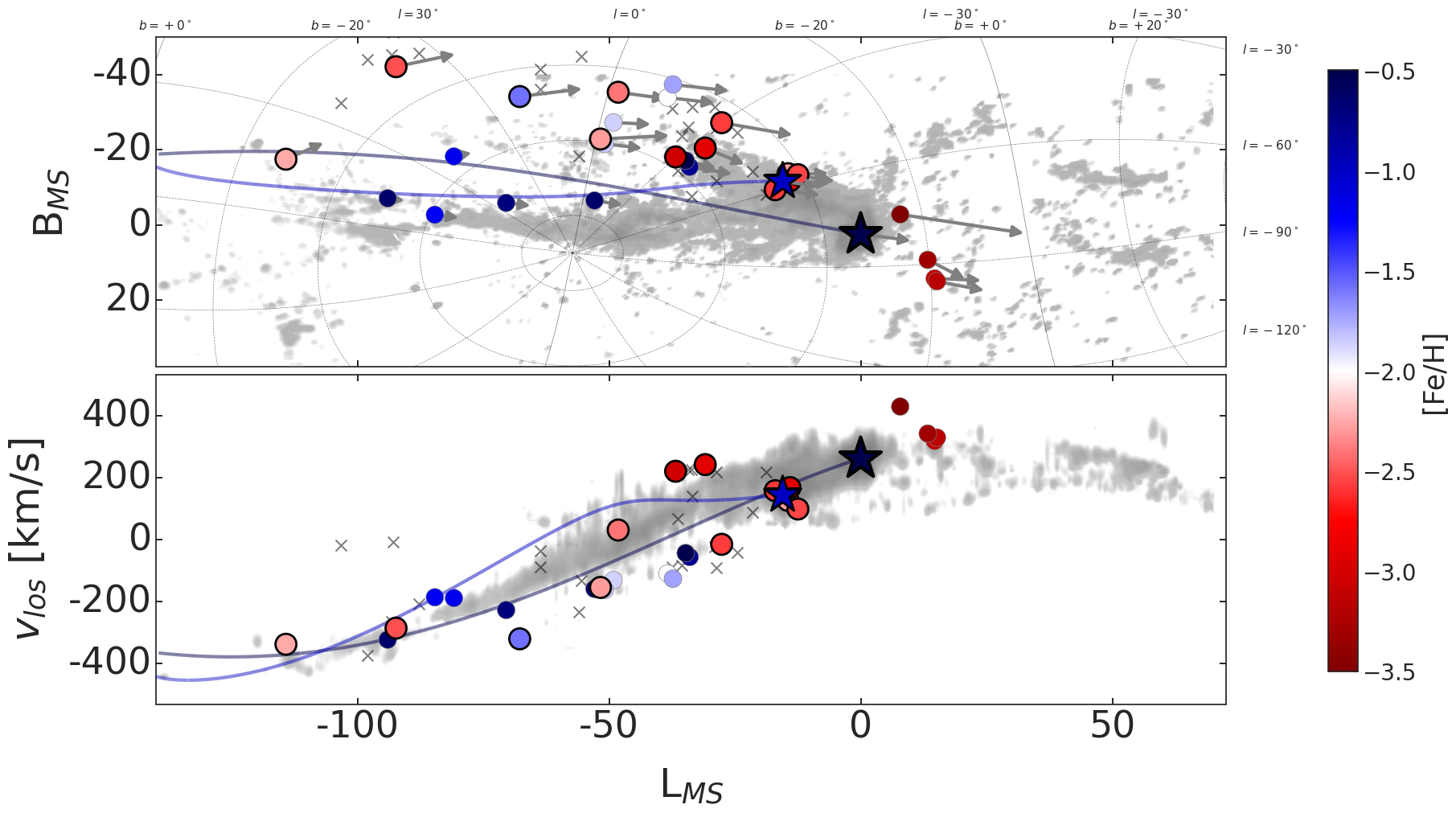}
   \caption{Comparison of the MSS stars in circles to the gas Stream from \citet{2010nidever} in gray density map in the Clouds aligned coordinates (top) and radial velocity (bottom). Literature stars from \citet{2023chandra}, \citet{2025placco}, and \citet{2025limberg,2026ji} are shown as small circles, the VMP MSS stars from this work are shown as large circles colour-coded by their spectroscopic metallicities, and the LMC/SMC themselves as large/small stars color-coded by their mean metallicities. Non-members from our spectroscopic follow-up are shown as black crosses. The quivers in the top panel indicate denote the proper motion of the MSS stars and the Clouds, scaled to 20 Myr of motion. The dashed lines trace the past 1 Gyr orbit of the Clouds \citep{2024vasiliev}. Note that most the leading arm of the gas Stream is at lower galactic latitudes where dust extinction is quite high.}
              \label{mss-magcoord}%
    \end{figure*}

Our final MSS candidate sample consists of 27 stars: the 12 newly confirmed GHOST members presented here and 15 candidates from recent literature \citep{2023chandra,2025placco,2025limberg,2026ji}. 
Their spatial, kinematic, and chemical properties are listed in Table~\ref{table1} and the spatial distribution of the gas MS, MSS stars, and the Clouds are shown in Galactic and MS-aligned coordinates in Figures~\ref{mss-sky} and \ref{mss-magcoord}.
Figure \ref{mss-magcoord} also shows the line-of-sight velocity distribution in the gas and the stellar stream of the Clouds.

A striking feature of the sample is the apparent metallicity-dependent spatial offset between the stellar and gaseous Streams. 
The more metal-rich stars broadly coincide with the H\,{\sc i} distribution, whereas the VMP and EMP candidates are displaced by approximately $20^\circ$ on the sky. 
This offset is seen in all of our VMP and EMP GHOST members and in the six metal-poor stars ([Fe/H]<-1.5) from \citet{2023chandra}. 
The close correspondence between the more metal-rich stars and the gas raises the possibility that some formed in situ from the stripped gas, as has been proposed for young stellar populations in the Bridge \citep{2017mackey}. 
However, this remains unconfirmed for the MSS, and is unlikely to explain the metal-rich debris identified by \citet{2023chandra}, whose analysis adopted old isochrones (4--10 Gyr).

Alternatively, the metal-rich and metal-poor stars may originate from different regions of the Clouds, stripped at different stages of their prolonged interaction.
Repeated interactions between the LMC and SMC could have removed populations with different initial metallicity distributions at different epochs, producing the spatial segregation observed today. 
In addition, the gaseous and stellar components need not follow the same spatial distribution: while tidal interactions can strip both components, hydrodynamical processes such as ram-pressure stripping can redistribute gas without directly affecting collisionless stars \citep{2013besla}. 
Offsets between stellar populations and H\,{\sc i} have also been observed in the Bridge \citep{2012casetti}. 
The VMP and EMP stars may therefore preferentially trace older populations in the outskirts of the SMC, rather than the more metal-rich populations associated with its star-forming regions. 
These processes could contribute to the observed offset, although they do not yet explain its full extent.
Despite their spatial offset from the gas, the proper motions of the MSS stars (Figure~\ref{mss-magcoord}, top panel) are aligned with the motion of the Clouds. 
Together with their large orbital energies and extreme negative $L_x$ values, this supports an association with the LMC.

The full MSS sample spans a wide metallicity range, from $[\mathrm{Fe/H}]=-0.5$ to $-4.3$, over $130^\circ$ on sky. 
Our 12 MSS stars include one intermediate metal-poor star ($[\mathrm{Fe/H}]=-1.6$), ten VMP stars, and one EMP star ($[\mathrm{Fe/H}]=-3.1$, MSS\_35200).
Together with the VMP and EMP candidates reported by \citet{2025placco}, \citet{2025limberg}, and \citet{2026ji}, our observations extend the known metal-poor stream members across both the leading and trailing sides of the LMC-SMC system.
On the leading side ($L_{\mathrm{MS}}>0^\circ$), the sample includes several EMP stars, including MSS\_11072 with $[\mathrm{Fe/H}]=-4.3$ \citep{2025limberg,2026ji}, demonstrating that the metal-poor population extends well beyond the main body of the Clouds.
Along the trailing stream ($L_{\mathrm{MS}}<0^\circ$), our GHOST sample complements the more distant and generally more metal-rich stellar debris identified by \citet{2023chandra}. 
As listed in Table~\ref{table1}, their sample probes heliocentric distances of approximately $60$--$100\,\mathrm{kpc}$ and predominantly higher metallicities ($[\mathrm{Fe/H}]\gtrsim-2$), whereas our GHOST stars lie at approximately $40$--$60\,\mathrm{kpc}$ and extend the known MSS candidate population into the VMP regime. 
This difference in distance is partly due to our magnitude-limited selection ($G\lesssim17.6$) from the Pristine-Gaia synthetic red-giant catalogue limiting our sensitivity to more distant (and metal-rich) giants.
The two samples therefore probe complementary regions of the stellar debris, and their different distance and metallicity distributions should be interpreted in the context of their respective selection functions.

The VMP and EMP candidates broadly follow the $v_{\mathrm{los}}$--$L_{\mathrm{MS}}$ trend of previously identified stellar debris and the gas Stream (Figure~\ref{mss-magcoord}, lower panel), but the non-members from our spectroscopic follow-up also exhibit a similar relation. 
Therefore, in Appendix~\ref{wake}, we compare the line-of-sight velocities of the MSS candidates with those of a Milky Way halo population in the same region of sky, and show that tangential velocities derived from distances and proper motions provide more effective discrimination between MSS members and general halo stars than line-of-sight velocities.

\subsection{Orbital histories}\label{sec:orbits}

\begin{figure*}
   \includegraphics[width=0.53\linewidth]{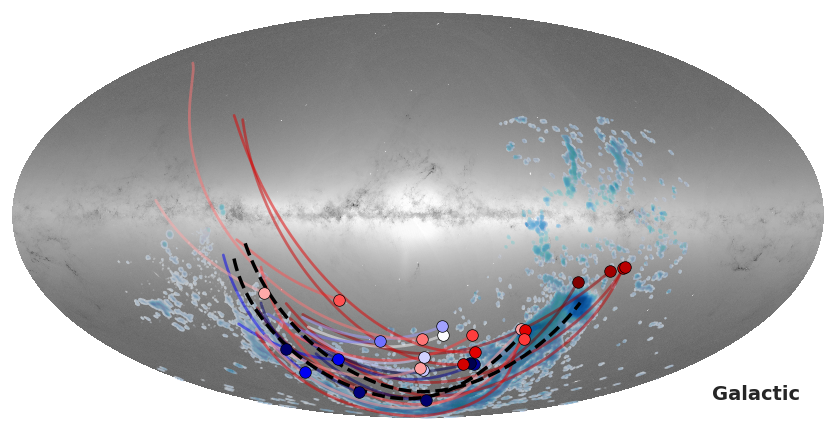}
   \includegraphics[width=0.47\linewidth]{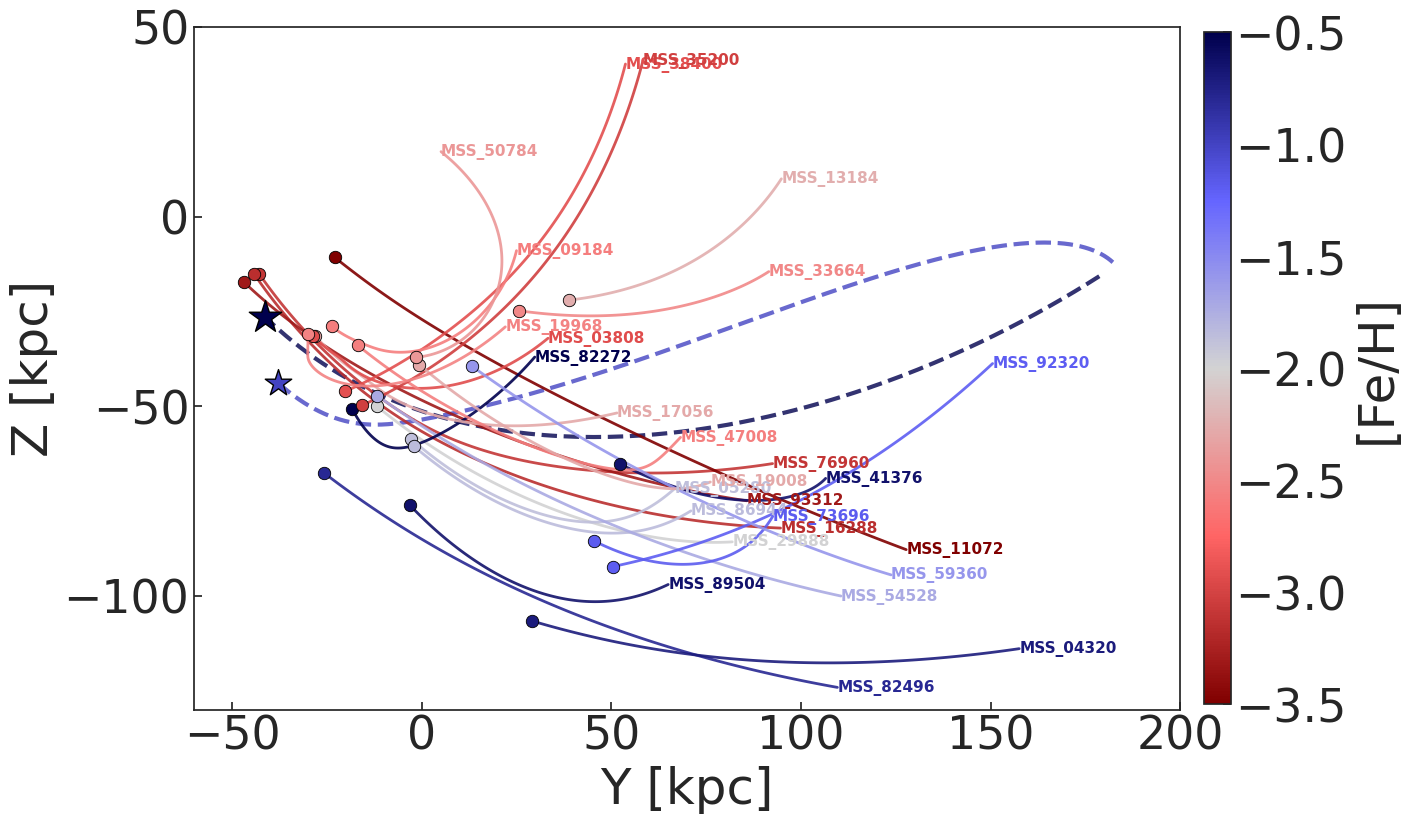}\\
   \includegraphics[width=0.52\linewidth]{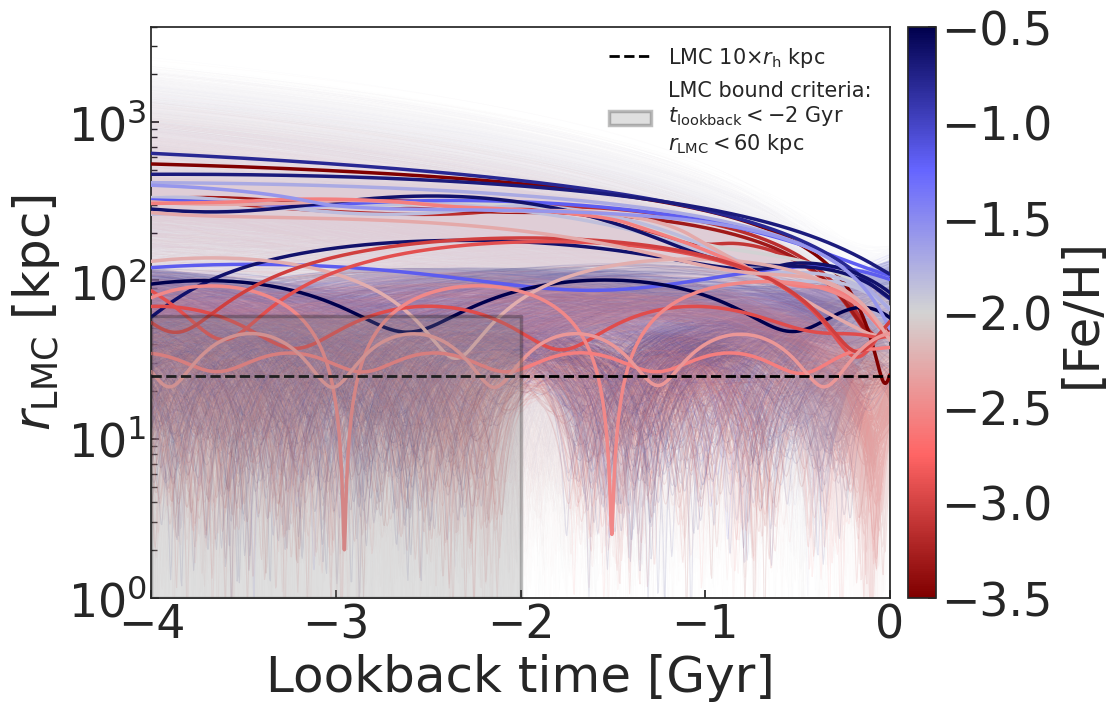}
   \includegraphics[width=0.48\linewidth]{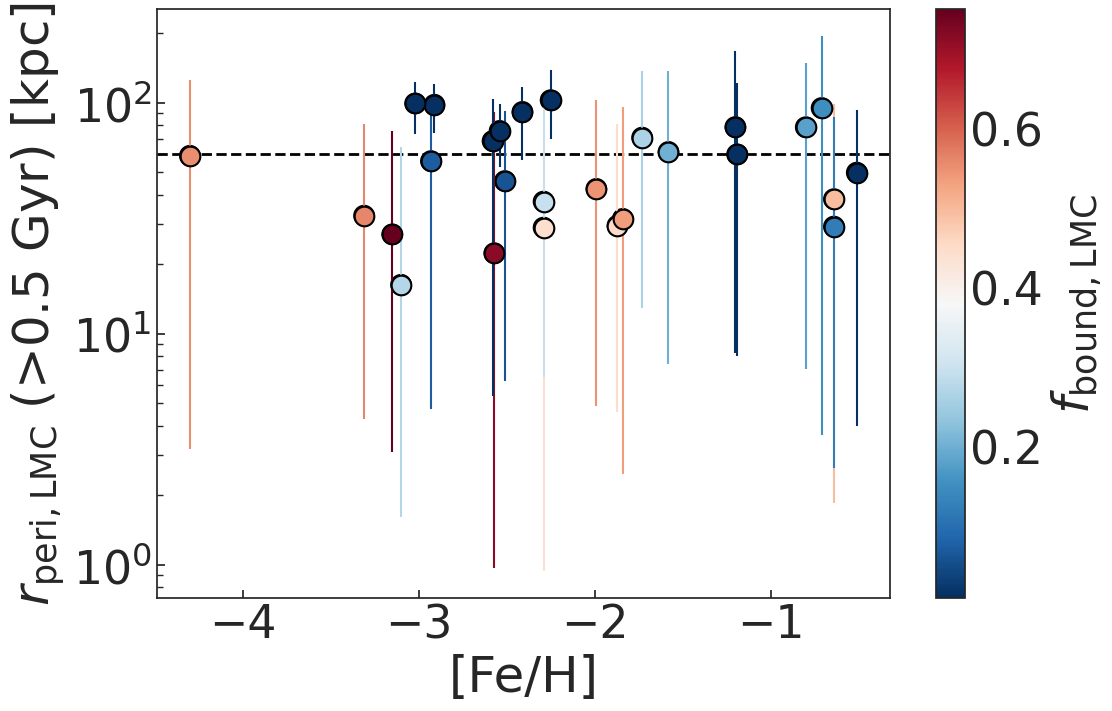}\\
   \caption{Orbital trajectories and LMC encounter metrics for the MSS stars as a function of metallicity. Top Left: Mollweide all-sky projection in Galactic coordinates showing the rewound mean orbits of the MSS stars (lines) and their present-day positions (filled circles), color-coded by metallicity $[\text{Fe/H}]$. Black dashed lines represent the mean rewound orbits of the LMC and SMC over a background Gaia density map with gas stream density contours overlaid in light blue. Top Right: Galactocentric $Y\text{--}Z$ Cartesian projection of the mean rewound orbits, with individual star identifiers labeled. Present-day positions of the Clouds are indicated by star symbols and their orbits are shown in dashed lines. Bottom Left: 3D distance relative to the LMC ($r_{\text{LMC}}$) versus lookback time for $1{,}000$ Monte Carlo (MC) orbital realizations per star and their mean orbit, color-coded by $[\text{Fe/H}]$. Bound/encounter realizations are highlighted in higher opacity, while unbound trajectories are rendered in lower opacity. The solid and dashed horizontal lines mark the $60\text{ kpc}$ encounter threshold and the LMC $10\times r_{\text{h}}$ radius ($\sim 25\text{ kpc}$), respectively. Bottom Right: Minimum LMC pericenter distance for lookback times older than $0.5\text{ Gyr}$ ($r_{\text{peri, LMC}}$) plotted against metallicity $[\text{Fe/H}]$. Error bars represent the $1\sigma$ percentile uncertainties derived from the MC realizations. Points are color-coded by the fraction of MC realizations that satisfy the LMC bound/encounter criteria ($f_{\text{bound, LMC}}$). The horizontal dashed line denotes the $60\text{ kpc}$ threshold.}
   \label{orbits}
\end{figure*}

Orbits provide an additional test of whether the past trajectories of the MSS stars are dynamically consistent with a Clouds' association. 
We therefore investigate the orbital histories of the 27 MSS stars. 

We perform Monte Carlo orbit integrations in the time-dependent Milky Way--LMC gravitational potential of \citet{2024vasiliev}, adopting their $\mathcal{L}2\mathcal{M}10first$ model, which represents a first-infall LMC with an initial halo mass of $2\times10^{11}\mathrm{M_\odot}$ orbiting within a Milky Way of virial mass $1\times10^{12}\mathrm{M_\odot}$. This model accounts for the evolving orbit and tidal mass loss of the LMC.
For each star, we generate $10^3$ realizations of its six-dimensional phase-space coordinates by sampling the observational uncertainties. 
Each realization is integrated backwards for 4 Gyr in the time-dependent Milky Way--LMC potential. 
At every timestep we compute the separation between the star and the LMC. 
Following the criterion introduced by \citet{2023chandra}, a realization is considered to have been bound to the LMC if it passes within 60 kpc of the LMC at lookback times greater than 2 Gyr, thereby excluding trivial recent encounters caused by the present-day proximity to the Clouds. 
The resulting $f_{\rm bound,LMC}$ quantifies the fraction of realizations consistent with a past LMC encounter under the adopted potential and criterion; it is not a probability of LMC membership.

Figure~\ref{orbits} shows the reconstructed orbital histories of the 27 MSS members. 
The top panels show representative median orbital trajectories over the past 1 Gyr, in Galactic coordinates and in the Galactocentric $Y$--$Z$ plane, illustrating that many stars follow trajectories consistent with the LMC in the past. 

The bottom left panel shows the evolution of the LMC--star separation for all Monte Carlo realizations, while the bottom right panel summarizes the corresponding bound fractions, $f_{\rm bound,LMC}$, for each star. 
Several stars have numerous realizations passing within the adopted 60 kpc threshold at lookback times greater than 2 Gyr, whereas others have substantially lower bound fractions.
The possibility of chance encounters with unrelated halo stars was investigated by \citet{2023chandra}, who found that only $\sim4\%$ of stars selected from a mock smooth halo satisfy the same criterion. Similarly, only one of our 21 GHOST non-members has a non-zero $f_{\rm bound,LMC}$ (below $10\%$), supporting its use as an additional dynamical diagnostic alongside our kinematic selection.
Overall, most MSS candidates have substantial bound fractions ($f_{\rm bound,LMC}\geq20\%$), with several exceeding 50\%. 
Nevertheless, seven stars (three from \citet{2023chandra} and four from this work) have $f_{\rm bound,LMC}\leq10\%$.  
As an independent check, we calculate the instantaneous orbital energy of each star with respect to the LMC and find a similar overall fraction of stars classified as bound to the LMC.
We retain the stars with low $f_{\rm bound,LMC}$ as MSS members because their orbital energy, angular momentum, and 6D positions and velocities remain consistent with an LMC origin, and thus the low bound fractions could be attributed to the stellar distance uncertainties and uncertainties in the MW--LMC potential.

We also repeat the integrations using the second-infall model of \citet{2024vasiliev}. 
Although the resulting bound fractions differ from those of our fiducial model, all 12 newly identified VMP MSS members retain non-zero $f_{\rm bound,LMC}$.
We further explore whether the MSS stars have closer orbital encounters with the SMC or any of the ten UFDs associated with the Clouds (Appendix~\ref{app:ufds}). We include a moving SMC model in the time-dependent MW--LMC potential, following \citet{2020patel}, and compare the minimum separations of each star from the LMC, SMC, and UFDs over lookback times of 2--4 Gyr.
This comparison does not reveal a clear preference for an association with any one of the systems, although some stars have smaller reconstructed pericentric distances to individual UFDs or the SMC than to the LMC. 
In particular, MSS\_47008 passes within $\sim2$ kpc of Reticulum~II. 
We also find tentative trends in the \citet{2025placco} stars which passes closer to Pictor~II and Reticulum~II, while the UMP star from \citet{2025limberg} and \citet{2026ji}, together with the \citet{2023chandra} stars at $\mathrm{[Fe/H]}<-1.5$, tend to pass closer to the SMC. 
However, given the uncertainties in the reconstructed stellar and UFD orbits and the adopted potential, these closest approaches alone cannot establish the host systems of these MSS stars.

The reconstructed orbits are sensitive to distance uncertainties, which propagate into the inferred tangential velocities and orbital energies. 
This is particularly relevant for RGB stars, whose distances depend on stellar models and metallicity estimates. We adopt a conservative 20\% distance uncertainty for both our sample and the literature stars. 
To assess its impact, we repeat the integrations for our 12 newly identified VMP MSS members using their individual posterior distance uncertainties. 
Four MSS stars then have $f_{\rm bound,LMC}=0$, which are the same four stars with $f_{\rm bound,LMC}<10\%$ under the uniform 20\% uncertainty. 
Nevertheless, these stars retain the extreme negative $L_x$ values and high orbital energies characteristic of the kinematically selected MSS population, and our main conclusions unchanged even if they are excluded from the MSS sample.

These orbital analyses provide dynamical consistency checks but cannot independently establish the origin of the VMP MSS population. 
Further constraints will be required from realistic simulations of the MW--LMC--SMC interaction, detailed chemical-abundance analyses, and additional members with reliable distance estimates, particularly standard-candle stars. 
As tangential velocities provide better discrimination between MSS members and the surrounding halo than line-of-sight velocities alone (Appendix~\ref{wake}), standard-candle stars in the MSS will be particularly valuable.

\subsection{Tidal evolution of an extended LMC stellar halo population}\label{lmchalo-test}

\begin{figure}
   \centering
   \includegraphics[width=\columnwidth]{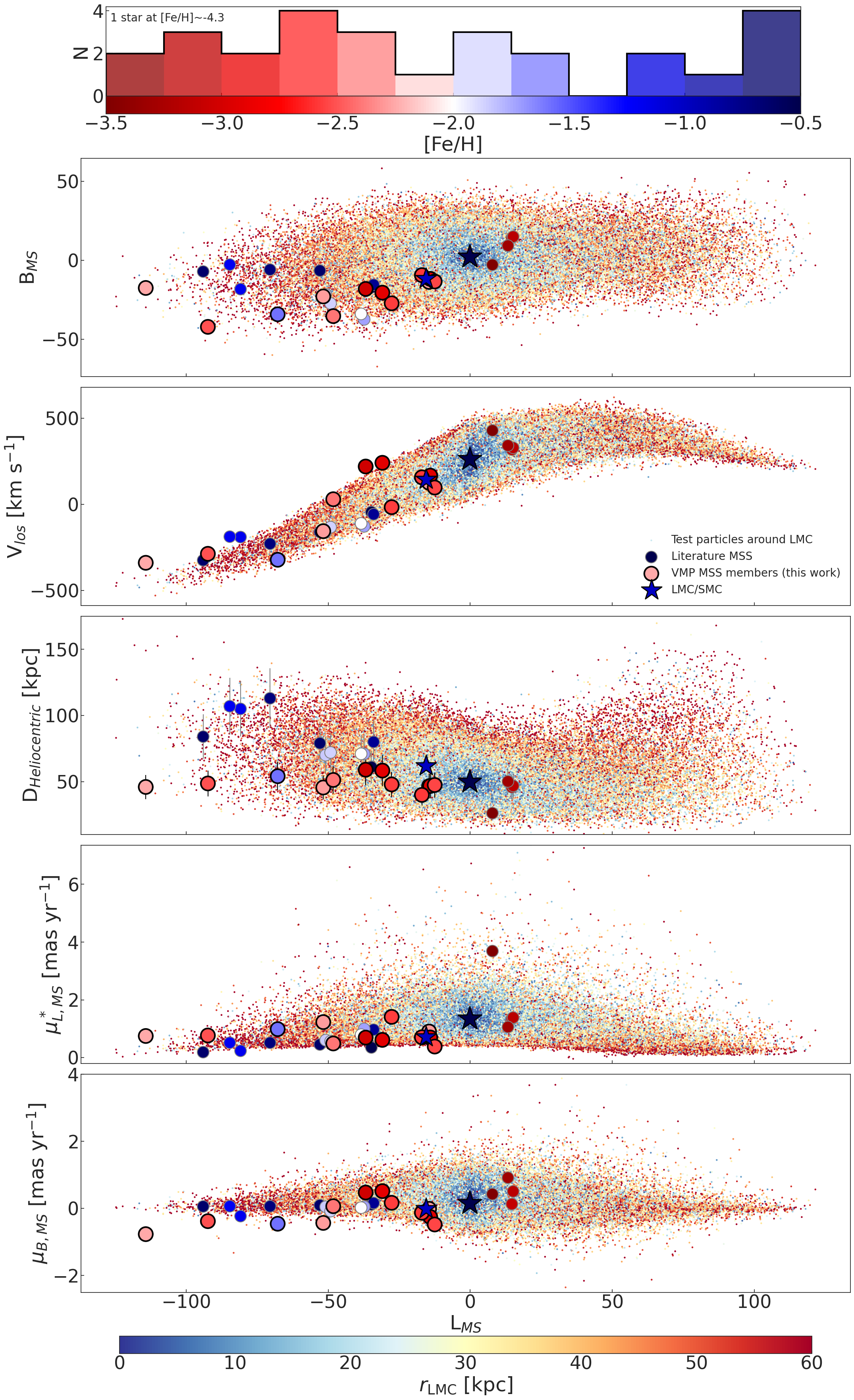}
   \caption{The MSS members from the literature (small circles) and this work (large circles) projected along the MS coordinates (panel 1), line-of-sight velocity (panel 2), heliocentric distances (panel 3), and proper motions along the transverse stream coordinates corrected for the solar-reflex motion (panels 4 and 5). The projection also includes test particles originating from a diffuse halo around the LMC integrated forward in a time-varying gravitational potential that includes both the Milky Way and a first-infall LMC from \citet{2021vasiliev}, LMC, SMC parameters adapted from \citet{2006kallivayalila,2006kallivayalilb}. The LMC, and SMC are colour-coded by their mean metallicities, and the MSS members are colour-coded by their spectroscopic metallicities. The histogram of the MSS members' metallicities are shown above the [Fe/H] colourbar.}
              \label{testparticles}%
    \end{figure}

To investigate whether the VMP MSS stars discovered in this work are compatible with a Clouds origin, we perform a complementary forward-modeling experiment in which a diffuse stellar population initially surrounding the LMC is evolved to the present day.
This is motivated by the fact that the most metal-poor MSS members appear to be systematically offset from both the gaseous MS and the more metal-rich stellar debris on the sky (subsection \ref{members}), and the distance uncertainties give rise to many different orbital configurations for our MSS stars (subsection \ref{sec:orbits}) . 

We generate a diffuse population of collisionless tracer particles initially surrounding the LMC (4 Gyr ago) and evolve them forward to the present day in the same time-dependent Milky Way--LMC potential \citep[$\mathcal{L}2\mathcal{M}10first$,][]{2024vasiliev}. 
We emphasize that this experiment is not intended to represent a physically motivated model for the stellar halo of the LMC. 
Instead, it serves only as a simple demonstration of where stars originating from an extended diffuse halo population around the LMC before infall would be expected to appear today under the adopted gravitational potential.

Initial conditions are generated using \texttt{Agama} \citep{2019vasiliev}. 
We assume a spherical, isotropic tracer population with a truncated power-law density profile,

\begin{equation}
\rho_{\rm halo}(r) \propto
r^{-2}\,
\exp\left[
-\left(\frac{r}{50~{\rm kpc}}\right)^3
\right].
\end{equation}

The particles are treated as massless tracers within the initial LMC potential. 
Their distribution function is obtained through Eddington inversion (Equation 4-140b of \citealt{1987binney}), allowing \texttt{Agama} to self-consistently sample particle positions and velocities via multidimensional rejection sampling. 
The resulting tracer population is then evolved to the present day using the same MW--LMC potential employed for the orbital integrations.

Figure~\ref{testparticles} compares the present-day distribution of these test particles with the observed MSS members identified in this work and the literature sample colour-coded by their spectroscopic metallicities. 
The particles are colour-coded according to their initial distance from the centre of the LMC, illustrating how stars originating at different radii populate distinct regions of phase space after tidal evolution. 
Despite the simplicity of the model, the simulated debris qualitatively reproduces the observed locations of both the 12 very metal-poor MSS members identified in this work and the 15 literature members remarkably well across all observable dimensions, including MS-aligned coordinates, line-of-sight velocity, heliocentric distance, and proper motions.

An important result is that the most metal-poor stars, which appear offset from both the H\,{\sc i} MS and the more metal-rich stellar debris on the sky, nevertheless occupy regions naturally populated by particles originating from an initially extended diffuse halo surrounding the LMC. 
This shows that the present-day phase-space distribution of the VMP MSS stars is consistent with an LMC association in the adopted gravitational potential.
However, while the model broadly reproduces the spatial extent of the observed debris, it predicts a population extending to $\sim40^\circ$ on both sides of the Stream in latitude, unlike the negative-latitude offset of the observed VMP stars.
This is because our experiment considers only a diffuse stellar population around the LMC and does not account for the dynamical interaction between the LMC and the SMC, which is thought to have contributed substantially to the formation of the tidal stellar debris, through the stripping of SMC material. 
More realistic models incorporating both Clouds, using test particles or self-consistent $N$-body simulations, will therefore be needed to investigate whether their interaction can explain the observed offset and its dependence on stellar metallicity. 
Such modelling is beyond the scope of the present work.

The MSS stars are also noticeably incomplete on the leading side of the Clouds, in comparison to the test particles. 
We searched for such a population within the red giant catalogue of \citet{2026viswanathan} but did not identify a statistically significant overdensity with distances and proper motions consistent with a Clouds origin. 
Interestingly, \citet{2022petersen} identified a population of RR Lyrae candidates in the leading arm region, although spectroscopic follow-up is still required to confirm their LMC association. 
A likely explanation for the limited number of RGB detections is observational incompleteness: much of the predicted leading arm lies at relatively low Galactic latitudes, where dust extinction and stellar crowding substantially reduce the efficiency of identifying distant, metal-poor halo giants. 
Future spectroscopic surveys such as 4MOST, and MOONS \citep{2019dejong, 2020gonzalez}, together with the improved astrometric precision and completeness expected from \textit{Gaia} DR4, will be helpful in identifying the putative leading stellar stream.

The particle distribution also reveals a dependence between the initial location of stripped particles and their present-day position in MS-aligned coordinates. 
Particles originating at larger initial distances from the LMC preferentially populate larger stream longitudes and latitudes, as expected for debris stripped from an extended stellar halo. 
However, given the simplified nature of the test-particle model, we refrain from interpreting these trends as detailed constraints on the stripping process. 
Future models that include additional physical ingredients, such as an initial metallicity gradient in the LMC stellar halo or the interaction and tidal stripping of different stellar populations of the SMC, will provide a more realistic framework for connecting the present-day chemical and kinematic properties of MSS stars to their initial pre-infall configuration.

Finally, we also consider the second-infall LMC model of \citet{2024vasiliev}, $\mathcal{L}2\mathcal{M}10$, which adopts the same underlying MW--LMC model as the first-infall $\mathcal{L}2\mathcal{M}10first$ potential used in this work, but the LMC is on its second infall into the Milky Way.
Repeating our test-particle integrations with this potential does not qualitatively change our conclusions: none of the observed MSS stars in our sample provide strong evidence either supporting or disfavoring a second-infall scenario for the Clouds.

\subsection{Origin of the very metal-poor MSS}

The metallicity distribution of the currently known MSS stellar population provides important clues to its origin. 
As shown in the top histogram panel of Figure~\ref{testparticles}, this work approximately doubles the number of confirmed MSS members, with the newly identified stars extending the population into the VMP regime. 
Although the bulk stellar populations of the LMC and SMC have higher mean metallicities ([Fe/H] $\sim-0.5$ and $\sim-1.0$, respectively), both Clouds host extended old and metal-poor populations. 
In particular, the fraction of metal-poor stars increases towards the outskirts of the SMC \citep{2008carrera,2014dobbiea,2014dobbieb,2016parisi,2023mucciarelli}, while spectroscopic studies have identified LMC stars extending into the VMP and EMP regimes, reaching [Fe/H] $\sim-4.2$ \citep{2008pompeia,2013swaelmen,2020nidever,2021mucciarelli,2021reggiani,2024chiti,2024oh}. 
Thus, although these stars represent only a small fraction of the Clouds' stellar populations, their presence in the MSS is not unexpected.

The SMC halo is a more plausible source of at least some of the VMP and EMP debris compared to the LMC halo, given its more metal-poor stellar population and an extended old component. 
Dynamical models predict multiple close encounters between the Clouds over the past several Gyr \citep{2006connors,2011diaz,2016besla,2026garver}, which could have stripped stars from different regions of the SMC at different epochs of interaction with the LMC. 
If these populations had distinct initial metallicity distributions, their subsequent tidal evolution could contribute to the spatial segregation observed between the metal-rich and metal-poor MSS stars. 

At present, it remains difficult to distinguish the origin of the VMP MSS stars using dynamics alone, and chemical abundances offer a potentially powerful additional diagnostic. 
However, abundance measurements of the most metal-poor stars in the Clouds are still based on relatively small and heterogeneous samples. 
While \citet{2021reggiani} found enhanced Eu abundances in VMP stars from both the Clouds, \citet{2024oh} reported broadly halo-like abundance patterns among their seven VMP and EMP LMC stars, without strong r-process enhancement. 
Carbon may provide another useful diagnostic: although CEMP fractions generally increase towards lower metallicities in the MW halo \citep{2014placco}, classical dwarf galaxies tend to exhibit lower fractions \citep{2024lucchesi,2024sestito,2024chiti}. 
The apparent deficit of CEMP stars in photometrically selected metal-poor LMC stars may, however, be partly due to selection biases against CEMP stars \citep{2019dacosta,2022arentsen,2024martin,2024chiti}. 
\citet{2026lucey} recently identified five CEMP-s stars around the LMC halo by specifically looking for carbon-enhanced stars in SDSS-V spectroscopy. 
Thus, although Eu, carbon, and other abundance ratios offer promising avenues for distinguishing the possible progenitors, no unique chemical signature of LMC or SMC origin has yet been established.

An interesting alternative hypothesis is that some VMP MSS stars originated from ultra-faint dwarf galaxies (UFDs) associated with the Clouds. 
The LMC has brought in its own population of satellites during its infall into the Milky Way \citep{2011sales,2016jethwa,2020patel,2021santossantos}, some of which could have been disrupted and contributed to the observed stellar debris. 
As discussed in Appendix~\ref{app:ufds}, the present-day configurations of known LMC-associated UFDs are broadly consistent with the regions occupied by our VMP MSS stars. 
Their chemical abundance patterns could provide an additional means of distinguishing this scenario from an origin in the Clouds themselves, as UFDs generally exhibit less extensive chemical enrichment and therefore, lower neutron-capture abundances than more massive systems such as the LMC and SMC \citep[e.g.,][]{2012venn,2014frebel,2018spite,2019ji,2020ji}. with an exception of a strongly r-process-enhanced UFD Reticulum~II \citep{2016ji}.
The [Ba/Mg], [Sr/Ba], and [Ba/Eu] ratios may help identify contributions from slow neutron-capture processes, particularly enrichment by metal-poor AGB stars, in addition to rapid neutron-capture enrichment \citep[e.g.,][]{2008pignatari,2014cescutti,2017mashonkina,2021cowan,2022mashonkina}. 
A contribution from disrupted globular clusters is also possible, although the metallicities of the known old LMC and SMC clusters go only down to [Fe/H]$\sim$--2.0, and --1.3, making this a less likely explanation for the origin of the VMP and EMP MSS stars \citep{2008mucciarelli,2010mucciarelli,2016dalessandro,2023mucciarellib}.

The high-resolution GHOST spectra obtained in this work will enable detailed abundance measurements to investigate these possible origins, with particular emphasis on light and neutron-capture elements. 
This analysis will be presented in a forthcoming paper.

\section{Conclusions and Outlook}\label{conclusions}

We have presented the first dedicated spectroscopic follow-up of VMP candidate members of the stellar Stream (MSS) associated with the LMC--SMC system, selected from the Pristine-Gaia synthetic red giants catalogue of \citet{2026viswanathan}. 

Our main results are summarised as follows:

\begin{itemize}

\item We obtained Gemini/GHOST spectroscopy for 33 candidate MSS stars and derived spectroscopic metallicities from the CaT using Voigt profile fitting. 
Twelve stars satisfy our kinematic membership criteria, including eleven VMP stars and one EMP star ([Fe/H] $\sim -3.0$). 
The remaining targets are associated with the nearby halo, the OC stream, or the MW outer halo dynamical wake. 
The 12 MSS members are accompanied by 13 MW halo/wake stars, yielding approximately 1:1 stream-to-wake ratio, consistent with the comparable contributions expected from N-body simulations of the MW-LMC interaction.

\item Combining our 12 new members with 15 previously identified stars approximately doubles the known sample of RGB stars associated with the MSS. 
The resulting 27 MSS stars spans more than 130$^\circ$ on-sky and a metallicity range of [Fe/H] $\sim -0.5$ to $-4.3$, demonstrating that the stellar component extends well into the VMP and EMP regimes.

\item The newly identified VMP stars exhibit a tentative on-sky offset of $\sim20^\circ$ from both the H\,{\sc i} MS and the previously identified, more metal-rich stellar debris, suggesting a metallicity-dependent spatial distribution. 
Despite this offset, their proper motions, radial velocities, and angular momenta are consistent with the orbital motion of the Clouds. 
Their distances of $\sim$40--60 kpc also extend the known VMP MSS population into an intermediate distance range between the leading-arm EMP stars and the more distant, metal-rich trailing-arm population.

\item Orbital integrations in a time-dependent MW--LMC potential show that many MSS stars have orbital histories consistent with a Clouds origin. 
Although the inferred LMC bound fractions are sensitive to stellar distance uncertainties and the adopted gravitational potential, the combined dynamical evidence supports an association with the Clouds. 
The test-particle simulation further demonstrate that a diffuse stellar population initially around the LMC can evolve into a distribution broadly consistent with the present-day phase-space locations of the observed MSS members, including the newly identified VMP stars. 
However, the model predicts a broad spatial distribution extending out to 40° on either side of the Stream, whereas the observed VMP stars are offset predominantly towards one side. 
This asymmetry may be related to the interaction with the SMC, which is not included in our simplified model, motivating more realistic simulations of the Clouds interaction.

\item The origin of the VMP population remains uncertain. 
These stars could have been stripped from the diffuse stellar halos of the LMC or SMC, or contributed by low-mass satellites associated with the Clouds. 
The present-day dynamical properties alone cannot distinguish between these possibilities, highlighting the importance of detailed chemical abundance measurements. 
The high-resolution GHOST spectra obtained here will enable such analyses, which will be presented in a forthcoming paper.

\end{itemize}

The newly identified VMP stars provide valuable observational constraints for future simulations of the MW–LMC–SMC interaction, combining their distances, 6D phase-space information, and metallicities to constrain the spatial and dynamical structure of the stellar debris and its possible progenitors.
These simulations will be essential to investigate the origin of the metallicity-dependent spatial offset and connect the observed debris to the initial structure of the Clouds. 
In parallel, upcoming wide-area spectroscopic surveys such as 4MOST \citep{2019dejong} and the improved astrometry expected from \textit{Gaia} DR4 will expand the census of the MSS, particularly its predicted leading arm, and help establish whether the different metallicity populations trace distinct wraps formed during multiple LMC--SMC interactions. 
Together, these efforts will provide new opportunities to constrain the assembly of the Clouds and their contribution to the outer Galactic halo.


\begin{acknowledgements}
AV thanks Ekta Patel, David Nidever, Pol Massana, Chervin Laporte, Adrian Price-Whelan, Guilherme Limberg, Vedant Chandra, Robin Chrisholm, and the \href{https://nfc.uvic.ca/}{\textcolor{black}{UVic NFC group}} for the helpful discussions about this work.
AV thanks Ewoud Wempe for the moral support.
AV acknowledges support from the Dutch Research Council (NWO) through a Rubicon grant, the Canadian Institute for Theoretical Astrophysics (CITA) through a CITA National Fellowship and the International Astronomical Union (IAU) and the Gruber Foundation (TGF) through a IAU TGF Fellowship.
KAV, AD, TC, JFN, SETS, JJ, DAB, AV, and DZ are grateful for funding through the National Science and Engineering Research Council Discovery Grants and CREATE programs.
FS acknowledges funding from the UK Science and Technology Facilities Council through grant ST/Y001443/1. 
The work of VMP is supported by NOIRLab, which is managed by the Association of Universities for Research in Astronomy (AURA) under a cooperative agreement with the U.S. National Science Foundation.
ES acknowledges funding through VIDI grant "Pushing Galactic Archaeology to its limits" (with project number VI.Vidi.193.093) which is funded by the Dutch Research Council (NWO). ES furthermore acknowledges funding from the European Research Council (ERC, EARLYMW, Grant number 101170507). Views and opinions expressed are those of the author(s) only and do not necessarily reflect those of the European Union or the European Research Council. Neither the European Union nor the granting authority can be held responsible for them.
Based on observations obtained at the international Gemini Observatory, a program of NSF NOIRLab, under programs GS-2025B-Q-310, GS-2023B-FT-104, GS-2024A-Q-221, and GS-2024A-Q-319, which is managed by the Association of Universities for Research in Astronomy (AURA) under a cooperative agreement with the U.S. National Science Foundation on behalf of the Gemini Observatory partnership: the U.S. National Science Foundation (United States), National Research Council (Canada), Agencia Nacional de Investigación y Desarrollo (Chile), Ministerio de Ciencia, Tecnología e Innovación (Argentina), Ministério da Ciência, Tecnologia, Inovações e Comunicações (Brazil), and Korea Astronomy and Space Science Institute (Republic of Korea).
This work has made use of data from the European Space Agency (ESA) mission Gaia (https://www.cosmos.esa.int/gaia), processed by the Gaia Data Processing and Analysis Consortium (DPAC, https://www.cosmos.esa.int/web/gaia/dpac/consortium). Funding for the DPAC has been provided by national institutions, in particular the institutions participating in the Gaia Multilateral Agreement.
AV acknowledges the use of large language models for language editing and coding assistance. All scientific analyses, interpretations, and conclusions are the authors' own.
AV also thanks the availability of the following packages and tools that made this work possible: astropy \citep{2022astropy}, NumPy \citep{2020numpy}, SciPy \citep{2020SciPy}, matplotlib (Hunter 2007), seaborn \citep{2021seaborn}, pandas \citep{2020pandas}, agama \citep{2019vasiliev}, gala \citep{2022pricewhelan}, healpy \citep{2019healpy}, JupyterLab \citep{2016jupyter}, iraf \citep{1986iraf}, pyraf \citep{2012pyraf}, dragons \citep{2023dragons}, astroquery \citep{2019astroquery}, ASPCAP \citep{2016aspcap}, doppler \citep{2015doppler}, emcee \citep{2013emcee}, corner \citep{2026corner}, PyAstronomy \citep{2019pya}, and Topcat \citep{2017topcat}. 
\end{acknowledgements}

%
\bibliographystyle{aa} 
\bibliography{biblio} 
%

\begin{appendix}
\section{Comparison of Pristine photometric metallicities with spectroscopic metallicities from GHOST}\label{app:photmet}

\begin{figure}
    \centering
    \includegraphics[width=\linewidth]{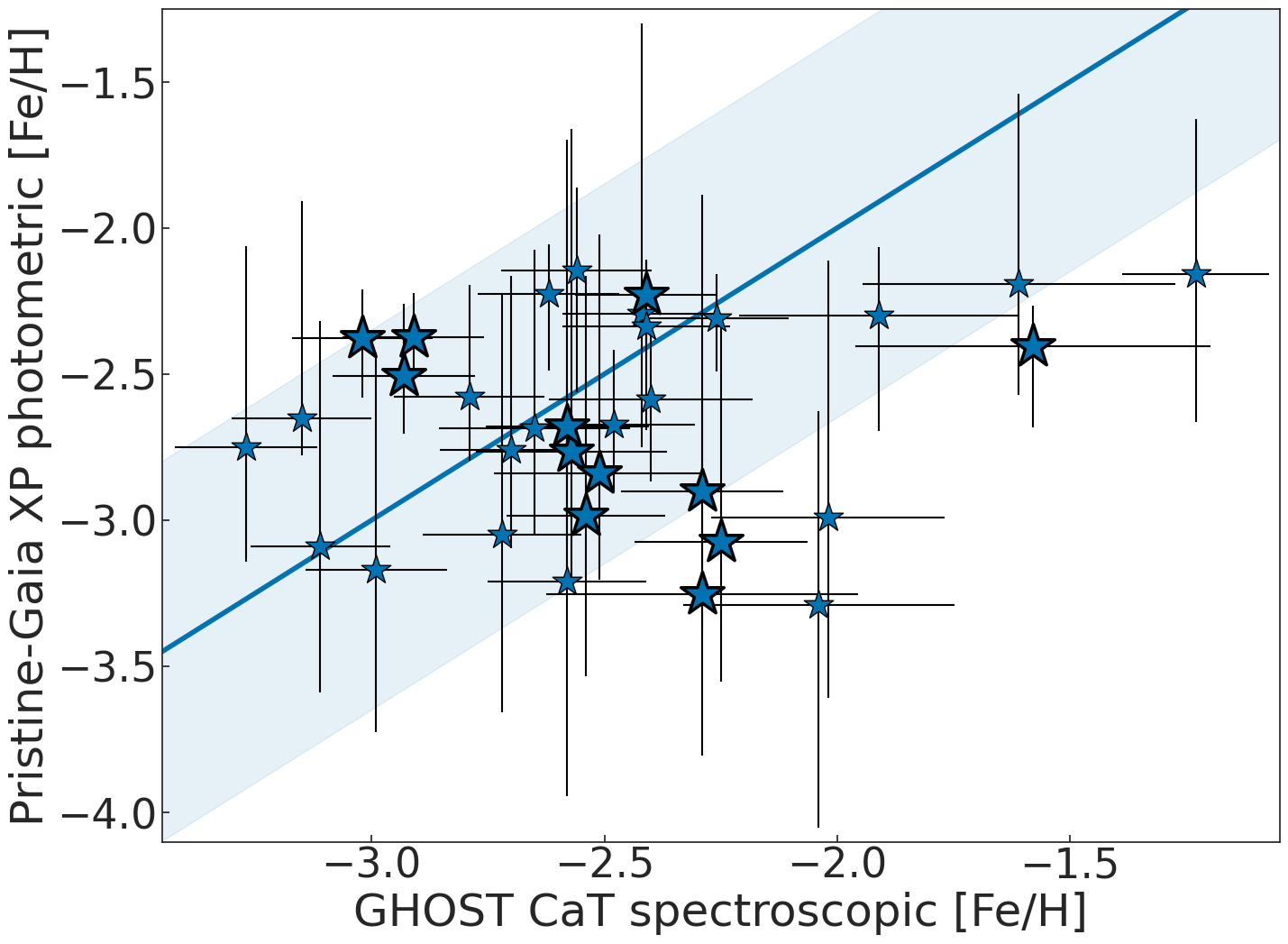}
    \caption{Comparison of Pristine-Gaia photometric metallicities with the CaT based spectroscopic metallicities inferred from the GHOST spectra. The one-to-one line and a 0.65 dex (median uncertainty of the photometric metallicities) offset around it is shown in blue. MSS member stars have larger symbols.}
    \label{fig:phot-met}
\end{figure}

Figure~\ref{fig:phot-met} compares the Pristine-Gaia photometric metallicities with the CaT-based spectroscopic metallicities derived from the GHOST spectra. 
Despite the relatively large uncertainties in the photometric metallicities, which are expected for these faint stars lying close to the magnitude limit of the Gaia DR3 XP spectra, the photometric and spectroscopic estimates show good overall agreement, with the differences generally comparable to the measurement uncertainties of the photometric metallicities. 
The median photometric metallicity uncertainty is $\sim0.65$ dex at these magnitudes. 
Of the 33 stars selected as VMP candidates, only four are found to have spectroscopic metallicities above [Fe/H] $= -2.0$, corresponding to a VMP selection success rate of $\sim87\%$. 
Three of these four stars remain close to the VMP boundary, while only one has [Fe/H] $> -1.5$. 
This demonstrates that the Pristine-Gaia selection remains highly effective at identifying VMP stars even at these faintest Gaia XP magnitudes probed by our sample. 
For comparison, \citet{2025viswanathan} reported a VMP selection success rate of $\sim$$96\%$ at brighter magnitudes with smaller CaHK uncertainties when applying the recommended quality cuts of \citet{2024martin}.



\section{Non-members from the GHOST MSS candidates}\label{app:nonmembers}

Of the 33 stars observed with GHOST, 21 are classified as non-members of the MSS. These stars fall into three categories based on their revised distances and their phase-space properties.
\textit{Nearby halo contaminants ($N = 4$):}
Our revised Bayesian photometric distances place four stars within $20\,\text{kpc}$ of the Sun, substantially closer than the Clouds and their expected stellar debris. Their total orbital energies are consequently more characteristic of the inner halo.
\textit{OC stream contaminants ($N = 4$):}
Four stars show spatial, proper-motion, and radial-velocity distributions consistent with the OC stream, suggesting that they belong to this distinct halo substructure rather than the MSS.
\textit{Background halo / wake stars ($N = 13$):}
The remaining 13 stars lie at larger distances and overlap with the background halo distribution. While their phase-space properties do not support a clear association with the MSS, their location along the trajectory is consistent with background halo stars in the region affected by the LMC's dynamical friction wake, the perturbation of the Galactic halo induced by the LMC's infall \citep{2019garavito,2021conroy}.
Notably, four of these 13 stars are EMP and lie beyond $55\,\mathrm{kpc}$, making them particularly interesting candidates for probing the early accretion history of the Milky Way and its low-mass progenitors. 
More broadly, these non-member stars from this work provides valuable high-resolution spectra for chemical abundance analysis: in addition to them being in the outer halo, most of the non-members are still VMP, including stars with a possible association to the OC stream.

The coexistence of stripped stellar debris and perturbed halo stars along this region is consistent with the dual-origin interpretation of the Pisces Plume \citep{2019belokurov,2021conroy,2023chandra}. 
In their sample of distant halo RGB stars, \citet{2023chandra} identified a lower limit of $\gtrsim 15\%$ of Pisces Plume stars as stripped debris from the Clouds, and suggested that this fraction could be as high as $\sim 50\%$. 
This higher fraction could help explain why the observed stellar overdensity is approximately twice as strong as predicted by the dynamical-response simulations of \citet{2021conroy}.
In our targeted sample of very metal-poor outer-halo stars, we identify 12 high-confidence MSS members and 13 background halo stars that may be associated with the wake, corresponding to a roughly $1:1$ ratio within our selection footprint and at these low metallicities. 
This ratio is suggestive but should not be interpreted as an estimate of the relative populations across the Pisces Plume, given our selection effects and small sample size. 
Whether it reflects spatial differences in the relative density of stream and wake stars, a dependence on metallicity, or both will require larger, homogeneous 6D phase-space and spectroscopic samples, and dedicated chemodynamical modelling.

\section{Observable properties of the final MSS members}\label{mss-table}

Table~\ref{table1} summarises the observed kinematic and dynamical properties of the 27 candidate MSS members classified in this work, including the 12 VMP stars identified here and the 15 candidates from the literature. 
For each star, we list its position and proper motions in MS coordinates, heliocentric distance, line-of-sight velocity, and metallicity, together with its orbital energy, $x$-component of the angular momentum, and the fraction of orbits bound to the LMC, $f_{\rm LMC}$, derived from the orbital analysis described in Section~\ref{orbits}. 
The adopted 6D positions, and velocities of LMC and SMC are also shown in the table \citep{2006kallivayalilb,2006kallivayalila}.

\begin{table*}
\centering
\caption{Kinematic and dynamical properties of MSS members and the Clouds (also available electronically on CDS)}
\label{table1}
\begin{tabular}{cccccccccccc}
\hline
\hline
Name & Gaia DR3 ID & L & B & $\mu_L^*$ & $\mu_B$ & d & v$_{los}$ & [Fe/H] & E/10$^{4}$ & L$_x/10^{3}$ & $f_{\rm LMC}$\\
 & & $\rm [deg]$ & $\rm [deg]$ & [mas & [mas & [kpc] & [km & & [km$^2$ & [kpc &in \%\\
 & & & & /yr] & /yr] & & /s] & & /s$^2$] & km/s] & \\
 \hline
\textbf{Limberg/Ji}\\
MSS\_11072 & 5262850721755411072 & 7.8 & -2.8 & 3.7 & 0.4 & 26.1 & 428.7 & -4.3 & 3.43 & -12.13 & 34.75 \\
\textbf{Placco}\\
MSS\_76960 & 5490389769744376960 & 14.7 & 14.3 & 1.3 & 0.1 & 45.7 & 317.9 & -3.1 & -2.62 & -14.05 & 80.5 \\
MSS\_16288 & 5491209253801516288 & 15.1 & 15.0 & 1.4 & 0.5 & 46.8 & 328.6 & -3.1 & -1.45 & -14.57 & 62.15 \\
MSS\_93312 & 5293243525250093312 & 13.3 & 9.3 & 1.1 & 0.9 & 50.1 & 341.5 & -3.3 & -0.91 & -12.46 & 61.2 \\
\textbf{Chandra}\\
MSS\_29888 & 6465479360347429888 & -38.4 & -33.9 & 0.8 & 0.0 & 71.0 & -110.6 & -2.0 & -1.29 & -14.67 & 60.75 \\
MSS\_05280 & 6568195388720305280 & -49.2 & -27.2 & 0.6 & -0.1 & 72.0 & -132.1 & -1.9 & -3.10 & -11.81 & 51.65 \\
MSS\_86944 & 6544712569129986944 & -50.9 & -21.5 & 0.6 & 0.1 & 70.0 & -163.3 & -1.8 & -2.48 & -13.09 & 59.75 \\
MSS\_54528 & 6477086664083654528 & -37.3 & -37.3 & 1.0 & 0.1 & 71.0 & -127.4 & -1.7 & 1.19 & -17.40 & 33.9 \\
MSS\_92320 & 2633051523542392320 & -84.6 & -2.7 & 0.5 & 0.1 & 107.0 & -187.2 & -1.2 & -0.56 & -29.57 & <10 \\
MSS\_73696 & 2601407505880573696 & -80.8 & -18.2 & 0.2 & -0.2 & 105.0 & -189.5 & -1.2 & -2.59 & -10.17 & <10 \\
MSS\_82496 & 6495806880338282496 & -34.0 & -15.4 & 1.0 & 0.2 & 80.0 & -57.2 & -0.8 & 2.72 & -28.71 & 26.1 \\
MSS\_04320 & 2392166457384504320 & -70.5 & -5.8 & 0.5 & 0.1 & 113.0 & -228.4 & -0.7 & 1.52 & -33.11 & 23.35 \\
MSS\_89504 & 2311053770911989504 & -52.9 & -6.5 & 0.5 & 0.1 & 79.0 & -160.1 & -0.6 & -2.53 & -14.48 & 55.85 \\
MSS\_41376 & 2661637550258741376 & -94.0 & -7.0 & 0.2 & 0.1 & 84.0 & -324.1 & -0.6 & -3.36 & -7.82 & 20.4 \\
MSS\_82272 & 6499025043497382272 & -34.8 & -17.1 & 0.4 & 0.1 & 61.0 & -44.4 & -0.5 & -5.26 & -6.68 & <10 \\
\textbf{Viswanathan}\\
MSS\_13184 & 1778940377006813184 & -114.2 & -17.4 & 0.7 & -0.8 & 45.9 & -339.0 & -2.2 & -4.03 & -6.08 & <10 \\
MSS\_17056 & 4685355019774317056 & -14.5 & -13.6 & 0.9 & -0.0 & 46.5 & 125.6 & -2.3 & -5.49 & -9.34 & 50.95 \\
MSS\_03808 & 4685608835179303808 & -14.1 & -11.7 & 0.7 & -0.3 & 46.7 & 167.0 & -2.9 & -6.34 & -7.14 & 15.75 \\
MSS\_47008 & 6408854408437947008 & -27.7 & -27.2 & 1.4 & 0.2 & 47.8 & -16.2 & -2.6 & -1.90 & -12.72 & 76.7 \\
MSS\_38400 & 6491278576058838400 & -30.9 & -20.4 & 0.6 & 0.5 & 58.3 & 241.3 & -2.9 & -3.02 & -9.95 & <10 \\
MSS\_59360 & 6810913576984559360 & -67.8 & -34.1 & 1.0 & -0.5 & 54.0 & -321.3 & -1.6 & 0.20 & -10.03 & 28.1 \\
MSS\_33664 & 6902720706255633664 & -92.3 & -42.1 & 0.8 & -0.4 & 48.6 & -286.8 & -2.5 & -4.70 & -5.60 & 14.65 \\
MSS\_09184 & 4690941913230909184 & -17.0 & -9.3 & 0.7 & -0.1 & 39.9 & 157.0 & -2.6 & -7.36 & -5.46 & 10.1 \\
MSS\_35200 & 6500075906032035200 & -36.8 & -18.1 & 0.7 & 0.5 & 59.0 & 219.8 & -3.0 & -2.73 & -11.53 & <10 \\
MSS\_19008 & 6545400279293019008 & -51.7 & -22.8 & 1.2 & -0.4 & 45.5 & -155.4 & -2.3 & -2.26 & -10.65 & 36.95 \\
MSS\_19968 & 4684339105387619968 & -12.5 & -13.3 & 0.4 & -0.5 & 47.4 & 97.8 & -2.5 & -6.63 & -4.03 & 10 \\
MSS\_50784 & 6578639928846550784 & -48.2 & -35.3 & 0.5 & 0.1 & 51.2 & 30.0 & -2.4 & -6.82 & -4.86 & <10 \\

\textbf{LMC} &  & -0.1 & 2.4 & 1.4 & 0.2 & 50 & 262.2 & -0.5 & -2.06 & -15.72 &  \\
\textbf{SMC} &  & -15.6 & -11.6 & 0.7 & -0.0 & 62 & 145.6 & -1.0 & -4.48 & -12.12 &  \\
\hline
\hline
\end{tabular}
\end{table*}

\section{The wake and the stream}\label{wake}

\begin{figure}
    \centering
    \includegraphics[width=\linewidth]{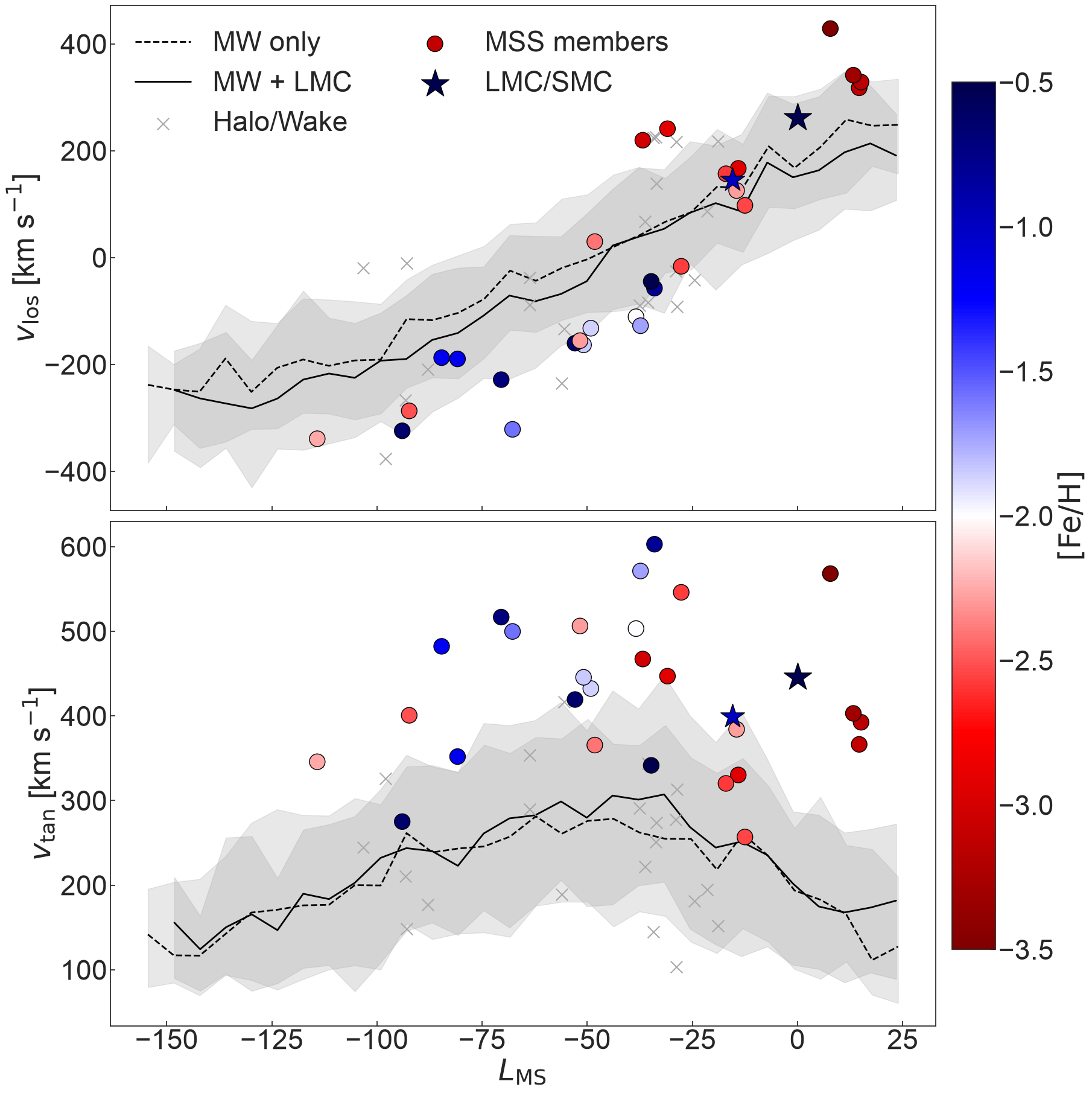}
    \caption{Kinematic distribution of MSS candidates, halo/wake stars, and model predictions as a function of MS-aligned longitude ($L_{\mathrm{MS}}$). Top panel: Line-of-sight velocity ($v_{\mathrm{los}}$) versus $L_{\mathrm{MS}}$. Bottom panel: Tangential velocity ($v_{\mathrm{tan}}$) versus $L_{\mathrm{MS}}$. Circles show candidate MSS members colored by metallicity ($\mathrm{[Fe/H]}$), blue star symbols denote the LMC/SMC, gray crosses shows MSS non-members followed-up in this work, and gray bands represent 1$\sigma$ synthetic halo test particles drawn from an \texttt{Agama} quasi-spherical distribution function and spatially cut to match the stream region ($40 \le d/\mathrm{kpc} \le 70$, $|l| \le 80^\circ$). Dashed and solid black lines show median tracks of the halo particles for MW-only and MW+LMC potentials from \citet{2021vasiliev}, respectively. While $v_{\mathrm{los}}$ overlaps significantly with the simulated halo background, $v_{\mathrm{tan}}$ cleanly isolates the stream members from the background halo/wake.}
    \label{fig:vtan_vlos_LMS}
\end{figure}

Disentangling MSS members from the underlying Milky Way stellar halo and the dynamically perturbed wake presents a significant observational challenge, because the MSS members are expected to occupy similar spatial footprint as the halo background that is influenced by the LMC's infall.

To construct a baseline for background halo contamination, we sampled mock halo test particles using \texttt{Agama} \citep{2019vasiliev} from a quasi-spherical distribution function in a spherical Milky Way potential. 
The synthetic particles were spatially filtered using the exact observational footprint of the MSS candidates as defined in \citet{2023chandra} and \citet{2026viswanathan}: Galactic longitudes $|l| \le 80^\circ$, heliocentric distances $40 \le d/\mathrm{kpc} \le 70$ (similar to the VMP MSS members identified in this work), and a latitude cuts defined by $b(l) = -60^\circ + 0.1l + 0.007l^2 \pm 20^\circ$.

Figure~\ref{fig:vtan_vlos_LMS} demonstrates the diagnostic power of full 6D phase-space information against the \texttt{Agama} halo test particles as a function of MS-aligned longitude ($L_{\mathrm{MS}}$). 
In the top panel ($v_{\mathrm{los}}$ vs. $L_{\mathrm{MS}}$), candidate MSS members (circles, colored by $[\mathrm{Fe/H}]$) closely track both the synthetic halo test particles (gray crosses) and the median model trajectories ($\mathrm{MW\ only}$ and $\mathrm{MW + LMC}$). 
Consequently, radial velocity alone fails to unequivocally separate MSS stars from background halo stars.

The bottom panel ($v_{\mathrm{tan}}$ vs $L_{\mathrm{MS}}$) reveals a more distinct kinematic separation. 
While the halo background remains tightly constrained at lower tangential velocities ($\lesssim 300\,\mathrm{km\,s^{-1}}$), the MSS members display higher transverse motions up to $v_{\mathrm{tan}}\sim350\text{--}600\,\mathrm{km\,s^{-1}}$. 
This pronounced separation in $v_{\mathrm{tan}}$ highlights the essential role of precise Gaia proper motions and spectrophotometric distance estimates ($v_{\mathrm{tan}} \propto d \cdot \mu$) in mapping this outer halo overdensity.

\section{The ultra-faint dwarf galaxies associated with the Clouds}\label{app:ufds}

\begin{figure}
    \centering
    \includegraphics[width=\linewidth]{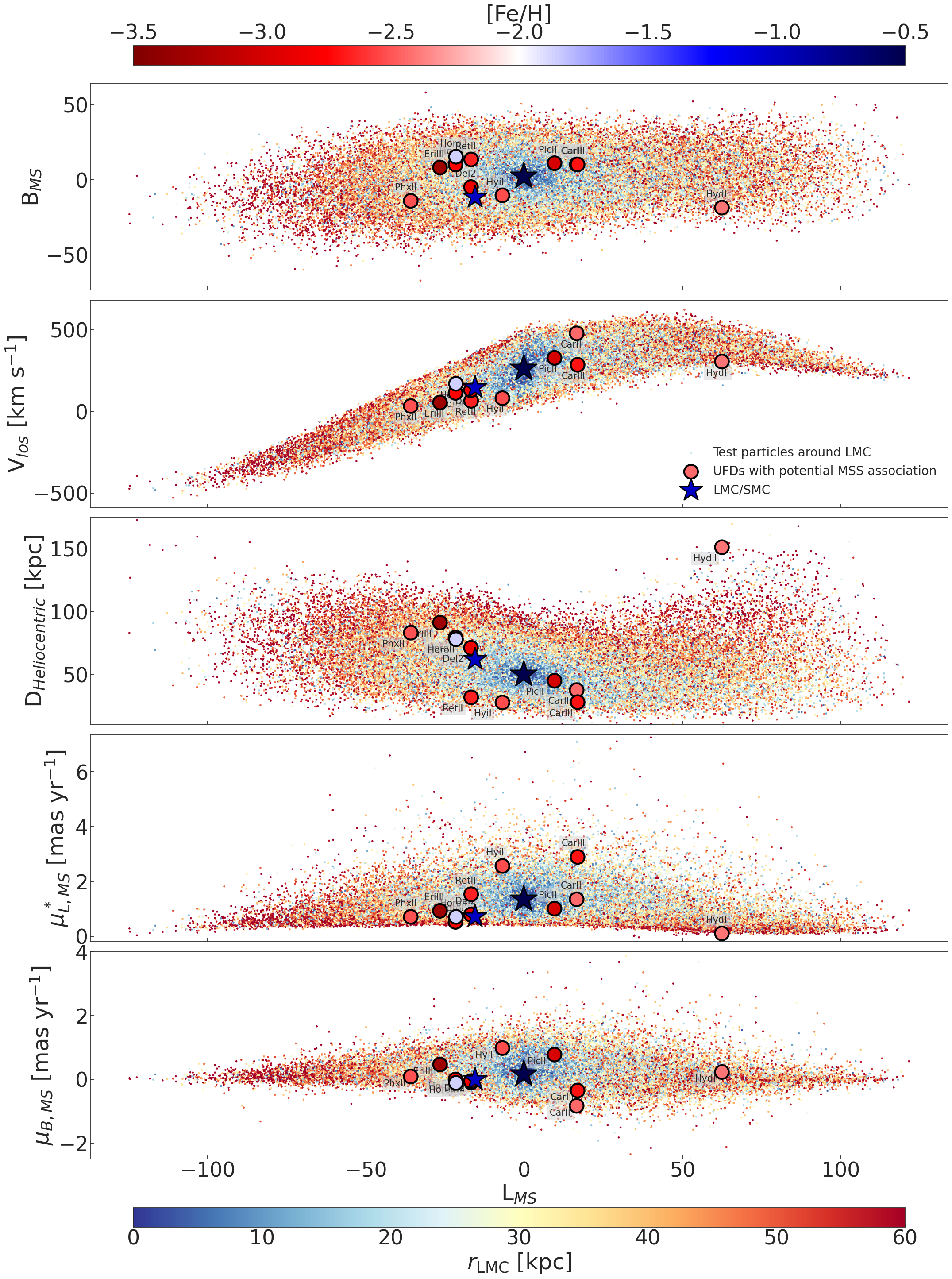}
    \caption{The 11 UFDs associated with the LMC (circles), together with the LMC and SMC, projected onto MS-aligned coordinates, line-of-sight velocity, heliocentric distance, and solar-reflex-corrected proper motions along the transverse stream coordinates. The diffuse LMC halo test particles are integrated forward in a time-varying Milky Way–LMC potential \citep{2021vasiliev}. The LMC, SMC, and UFDs are colour-coded by their mean spectroscopic metallicities \citep{2025pace}. Hydra~II, whose observed properties closely match the simulated debris despite its uncertain LMC association, is included as a candidate former LMC satellite.}
    \label{fig:ufds}
\end{figure}

\begin{figure*}
    \centering
    \includegraphics[width=\linewidth]{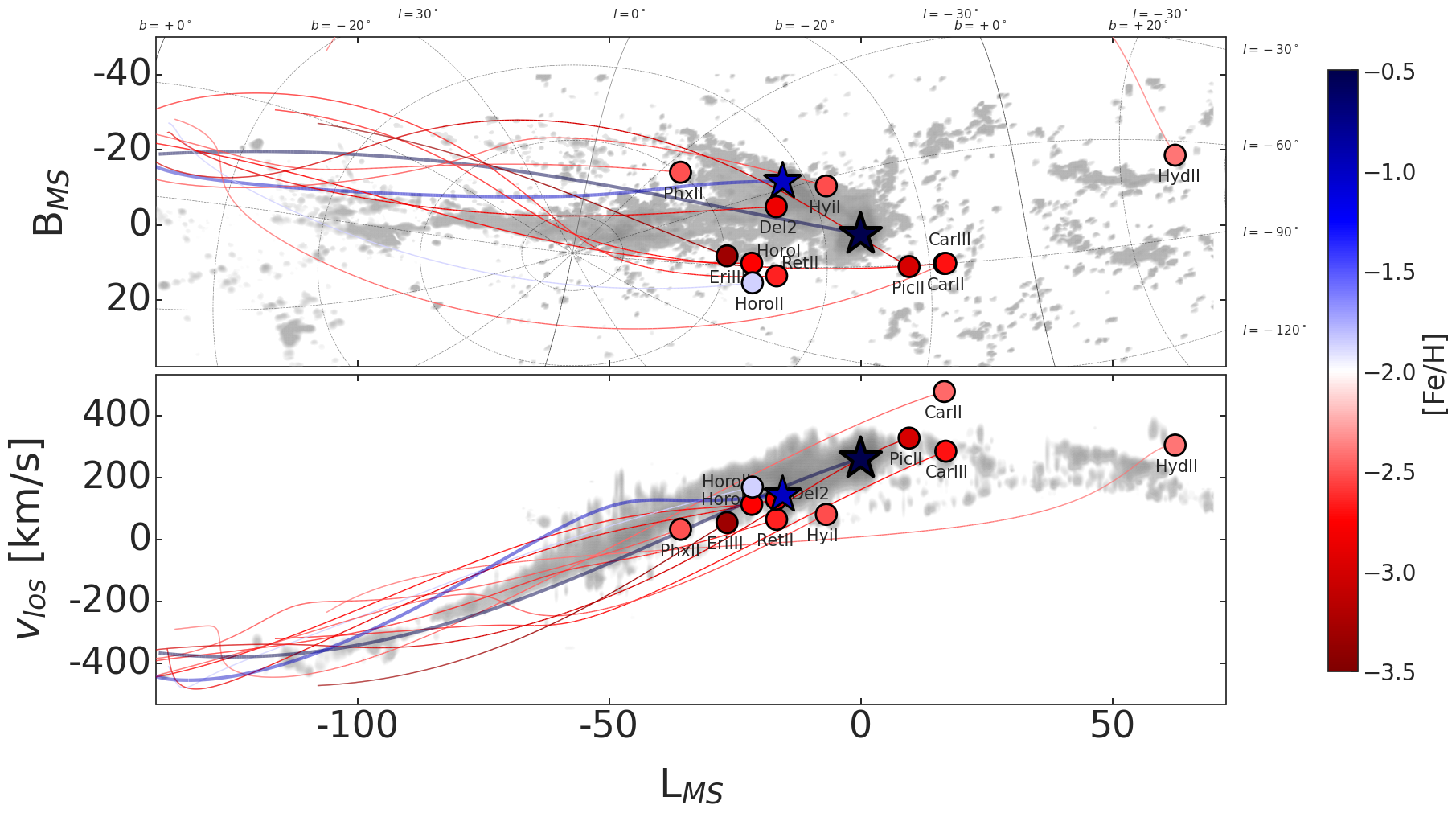}
    \caption{The 11 UFDs associated with the Clouds, together with the LMC and SMC, shown in MS-aligned coordinates (top) and MS-aligned longitude versus radial velocity (bottom), with the gas Stream from \citet{2010nidever} shown as a grey density map in the background. The LMC and SMC are represented by large star symbols, while the 11 UFDs are shown as circles; all are colour-coded according to their mean metallicities. The corresponding orbits of the LMC, SMC, and UFDs, integrated in the time-varying Milky Way+first-infall LMC potential from \citet{2024vasiliev}, are also shown. This figure is directly comparable to the MSS stars shown in Figure~\ref{mss-magcoord}.}
    \label{fig:ufds-orbit}
\end{figure*}

As a massive first-infall satellite, the LMC brought a group of ultra-faint dwarf galaxies (UFDs) into the Milky Way system, where they are now subject to tidal perturbations within the host halo. 
Figure~\ref{fig:ufds} compares the observed properties of 11 UFDs considered in this work, together with the LMC and SMC, against the simulated diffuse LMC stellar halo integrated forward in a time-varying Milky Way plus first-infall LMC potential \citep{2024vasiliev}. 
Our sample comprises ten UFDs with high-probability LMC associations ($P_{\rm LMC}>90\%$) according to \citep[$\mathcal{L}2\mathcal{M}10first$ MW+LMC time-varying potential,][]{2024vasiliev}, and Hydra~II, which we hypothesize to be a Clouds-associated UFD. 
We do not include Carina~dSph, because its association with the LMC remains debated and it is also not an ultra-faint dwarf. 
For the mean observed phase-space properties, we use the Local Volume Database compilation of \citet{2025pace}. 
Where line-of-sight velocities or spectroscopic metallicities are unavailable in that catalogue, we supplement the data with measurements from the literature: the line-of-sight velocity and metallicity of Horologium~II from \citet{2019fritz}, and those of Delve~2 from \citet{2026cerny}; for Carina~III, we adopt the mean spectroscopic metallicity from \citet{2020ji}.

Across the five diagnostic projections---MS-aligned coordinates ($L_{\mathrm{MS}}, B_{\mathrm{MS}}$), line-of-sight velocities ($v_{\mathrm{los}}$), heliocentric distances ($d_{\odot}$), and solar-reflex-corrected proper-motion components---the ten high-probability LMC satellites closely follow the coherent tidal structures traced by the simulated LMC test particles. 
Hydra~II, despite having a substantially lower LMC-association probability in \citet{2024vasiliev} ($P_{\rm LMC}\sim 40\%$), also exhibits a striking correspondence with the simulated debris in most of the observed phase-space dimensions. 
In particular, its position and radial velocity are consistent exactly with those of the gas Stream. 
Its present location in the Leading Arm, at a considerable distance from the current position of the LMC, could be consistent with an origin as a loosely bound outer-halo LMC satellite that was stripped during the earliest stages of the LMC's infall into the Milky Way halo. 
We therefore include Hydra~II as a candidate former LMC satellite, while recognising that its association remains uncertain.

The phase-space distribution of LMC-associated UFDs also motivates a possible connection between the disruption of LMC satellites and the very metal-poor (VMP) members of the MSS identified in this work. 
UFDs are generally more metal-poor than the LMC and SMC, whose mean metallicities are substantially higher. 
Tidal disruption of UFDs accreted along with the LMC could therefore contribute a population of VMP stars to the MSS. 
This provides a plausible origin scenario for the VMP MSS members, which constitute approximately half of the known MSS population to date.

\section{List of MSS candidates and their Gemini/GHOST observations}

Table~\ref{appendixtable1} provides the Gemini/GHOST observation log for the 33 candidates observed in this work, including the exposure times, achieved signal-to-noise ratios, and radial-velocity measurements per exposure. 
Table~\ref{appendixtable2} lists the 41 MSS candidates identified in \citet{2026viswanathan}, together with their Gaia DR3 astrometry, Pristine-Gaia photometric metallicities, distances, radial velocities, and, where available, CaT-based spectroscopic metallicities.

\begin{table*}
\centering
\caption{Observed MSS candidates (also available electronically on CDS). The classification of MSS members, halo/wake stars, OC stream, and nearby giant stars are based on section \ref{e-lz-selection}, and Appendix \ref{app:nonmembers}}\label{appendixtable1}
\begin{tabular}{ccccccccccc}
\hline
\hline
Gaia DR3 Source ID &
RA &
Dec &
$\mu_{\alpha}^*$ &
$\mu_{\delta}$ &
Gaia $G$ & 
[Fe/H]$_{\rm PGS}$ & 
d & 
v$_{los}$ &
[Fe/H]$_{\rm CaT}$ \\
 & [deg] & [deg] & [mas & [mas &  &  & [kpc] & [km/s] & \\
  &  &  & /yr] & /yr] &  &  &  &  & \\
\hline
\textbf{MSS members}\\
1778940377006813184 & 328.899 & 17.201 & -0.16 & -1.58 & 16.68 & -3.07 & 45.92 & $-339.0\pm0.1$ & $-2.25\pm0.19$ \\
4684339105387619968 & 11.196 & -76.222 & 0.32 & -1.10 & 16.52 & -2.98 & 47.40 & $97.8\pm0.1$ & $-2.54\pm0.17$ \\
4685355019774317056 & 7.734 & -74.485 & 0.88 & -1.50 & 16.84 & -2.90 & 46.45 & $125.6\pm0.1$ & $-2.29\pm0.17$ \\
4685608835179303808 & 14.836 & -74.165 & 0.72 & -1.30 & 16.05 & -2.51 & 46.68 & $167.0\pm0.2$ & $-2.93\pm0.15$ \\
6408854408437947008 & 329.614 & -61.959 & 0.05 & -2.41 & 17.17 & -2.77 & 47.83 & $-16.2\pm0.0$ & $-2.57\pm0.21$ \\
6491278576058838400 & 344.079 & -59.734 & 0.83 & -1.39 & 16.02 & -2.37 & 58.30 & $241.3\pm0.3$ & $-2.91\pm0.15$ \\
6810913576984559360 & 324.460 & -27.153 & -0.26 & -1.93 & 15.96 & -2.40 & 54.04 & $-321.3\pm0.2$ & $-1.58\pm0.38$ \\
6902720706255633664 & 312.985 & -9.386 & -0.16 & -1.73 & 16.86 & -2.84 & 48.58 & $-286.8\pm0.2$ & $-2.51\pm0.23$ \\
4690941913230909184 & 17.648 & -70.540 & 1.01 & -1.36 & 16.83 & -2.68 & 39.91 & $157.0\pm0.1$ & $-2.58\pm0.17$ \\
6578639928846550784 & 322.256 & -43.149 & 0.10 & -1.50 & 15.73 & -2.23 & 51.22 & $30.0\pm0.1$ & $-2.41\pm0.15$ \\
6500075906032035200 & 347.854 & -54.064 & 0.88 & -1.42 & 16.21 & -2.38 & 58.95 & $219.8\pm0.5$ & $-3.02\pm0.15$ \\
6545400279293019008 & 339.128 & -40.505 & 0.01 & -2.35 & 17.17 & -3.25 & 45.49 & $-155.4\pm0.1$ & $-2.29\pm0.34$ \\

\textbf{MW halo/wake}\\
6903270668228195840 & 309.732 & -9.562 & -0.14 & -0.92 & 15.63 & -2.31 & 47.88 & $-267.0\pm0.1$ & $-2.26\pm0.16$ \\
4923149526247840640 & 6.513 & -54.750 & 0.28 & -0.90 & 15.89 & -2.65 & 56.44 & $226.0\pm0.1$ & $-3.15\pm0.15$ \\
4706631669280837248 & 0.207 & -68.058 & 0.73 & -0.63 & 16.54 & -2.75 & 42.60 & $86.9\pm0.1$ & $-3.27\pm0.15$ \\
4905669009352040448 & 2.260 & -60.626 & -0.00 & -1.54 & 16.95 & -3.21 & 43.04 & $-91.4\pm0.1$ & $-2.58\pm0.17$ \\
6908181396394873600 & 309.986 & -6.006 & -0.71 & -1.34 & 17.04 & -2.99 & 45.33 & $-376.0\pm0.1$ & $-2.02\pm0.25$ \\
6788174100271623424 & 316.226 & -30.467 & 0.38 & -1.20 & 16.65 & -2.76 & 48.57 & $-37.2\pm0.2$ & $-2.70\pm0.15$ \\
6498000990151085696 & 354.195 & -54.006 & 0.30 & -0.72 & 16.02 & -3.09 & 59.89 & $67.2\pm0.1$ & $-3.11\pm0.15$ \\
6507990809203290368 & 333.817 & -56.349 & -0.14 & -0.67 & 16.64 & -3.17 & 44.73 & $225.0\pm0.1$ & $-2.99\pm0.15$ \\
6554181430124375936 & 344.119 & -35.963 & 0.18 & -0.70 & 16.40 & -2.58 & 55.24 & $-235.0\pm0.1$ & $-2.79\pm0.16$ \\
6452441557622974336 & 322.212 & -59.649 & 0.26 & -0.94 & 16.71 & -2.59 & 60.04 & $-25.4\pm0.1$ & $-2.40\pm0.22$ \\
6460034510046368000 & 324.308 & -55.963 & 0.09 & -1.17 & 16.80 & -2.69 & 49.13 & $138.3\pm0.1$ & $-2.65\pm0.21$ \\
6398469211875052544 & 333.887 & -65.256 & 0.43 & -0.75 & 16.94 & -3.05 & 44.15 & $-41.6\pm0.2$ & $-2.72\pm0.17$ \\
2692140476715314304 & 319.223 & 2.028 & 0.21 & -1.05 & 16.29 & -2.67 & 48.11 & $-19.8\pm0.2$ & $-2.48\pm0.17$ \\

\textbf{OC stream}\\
6462341629039072896 & 326.157 & -52.665 & 0.23 & -1.05 & 15.94 & -2.23 & 57.13 & $-89.8\pm0.7$ & $-2.62\pm0.15$ \\
6505337859444498688 & 338.215 & -55.351 & 0.82 & -1.85 & 16.56 & -2.34 & 35.74 & $-83.5\pm0.1$ & $-2.41\pm0.18$ \\
6779255411703094656 & 311.175 & -35.940 & -0.47 & -2.38 & 16.77 & -2.15 & 36.21 & $-133.4\pm0.2$ & $-2.56\pm0.16$ \\
6785860826587863040 & 322.377 & -30.614 & 0.39 & -1.64 & 16.68 & -3.29 & 44.35 & $-87.9\pm0.2$ & $-2.04\pm0.29$ \\
\textbf{Nearby giants}\\
6880996005762433664 & 303.804 & -11.399 & 0.37 & -2.39 & 16.65 & -2.19 & 12.94 & $-9.8\pm0.2$ & $-1.61\pm0.34$ \\
6406319587459014912 & 332.670 & -61.226 & 0.67 & -1.03 & 16.79 & -2.30 & 17.71 & $216.5\pm0.1$ & $-1.91\pm0.30$ \\
6899708353634042752 & 310.264 & -13.362 & 1.13 & -1.56 & 17.09 & -2.29 & 19.36 & $-208.9\pm0.1$ & $-2.42\pm0.17$ \\
4692493079322662528 & 18.522 & -68.277 & 1.47 & -1.15 & 17.22 & -2.16 & 17.16 & $218.1\pm0.2$ & $-1.23\pm0.16$ \\
\textbf{MSS candidates}\\
1762999382550564224 & 311.701 & 16.090 & 0.140 & -0.796 & 16.22 & -2.70 & 47.6 \\
1854485759380322944 & 318.994 & 32.478 & 0.294 & -0.261 & 16.14 & -2.64 & 46.9 \\
2701901700428106880 & 325.226 & 8.079 & -0.092 & -0.984 & 16.39 & -3.03 & 43.1 \\
6480458075971978240 & 317.508 & -46.534 & -0.185 & -1.633 & 16.71 & -3.00 & 52.6 \\
6507867148505055616 & 335.537 & -56.163 & 0.977 & -1.765 & 16.82 & -2.70 & 40.0 \\
6521341835302643072 & 358.203 & -53.128 & 0.058 & -1.357 & 17.09 & -2.49 & 47.7 \\
6563968732880189184 & 324.131 & -45.482 & -0.010 & -0.869 & 16.30 & -2.78 & 50.1 \\
6816729856056368128 & 326.348 & -22.461 & 0.862 & -1.328 & 16.69 & -3.16 & 48.8 \\
\hline
\hline
\end{tabular}
\end{table*}

\onecolumn
{
\setlength{\tabcolsep}{3.5pt}
\begin{longtable}{
p{2.7cm}
p{2.8cm}
p{3.5cm}
p{2.1cm}
p{1.43cm}
p{2.06cm}
p{1.cm}
p{1.cm}
}
\caption{Gemini/GHOST observations of candidate MSS members.}\\
\hline
\hline
Program & File Name & Gaia DR3 ID & Object & Blue/Red & v$_{los}$ per & S/N & Gaia $G$\\&&&&Exposure times (s)&Exposure (km/s)&at CaT &\\
\hline
\endfirsthead
\hline
Program & File Name & Gaia DR3 ID & Object & Blue/Red & v$_{los}$ per & S/N & Gaia $G$\\&&&&Exposure times (s)&Exposure (km/s)&at CaT &\\
\hline
\endhead
\multicolumn{1}{c}{\textbf{MSS members}}\\
GS-2024A-Q-319 & S20240530S0096 &1778940377006813184& MSS\_13184 & 1 x 1100 & -338.95 ± 0.08 & \multicolumn{1}{c}{36.2} & 16.68 \\
GS-2023B-FT-104 & S20231208S0062&4684339105387619968 & MSS\_19968 & 1 x 800 & 97.78 ± 0.14 & \multicolumn{1}{c}{25.8} & 16.52\\
GS-2025B-Q-310 GS-2023B-FT-104 & S20251119S0040 S20251119S0041 S20251027S0053 S20250930S0064 S20231208S0061 &4685355019774317056& MSS\_17056 & 1 x 1300 1 x 1300 1 x 1300 1 x 1300 1 x 1200 & 125.56 ± 0.10 125.57 ± 0.07 125.61 ± 0.19 125.53 ± 0.03 125.59 ± 0.07 & \multicolumn{1}{c}{43.2} & 16.84\\
GS-2025B-Q-310 & S20250913S0062&4685608835179303808 & MSS\_03808 & 1 x 900 & 166.99 ± 0.19 & \multicolumn{1}{c}{21.8} & 16.05\\
GS-2024A-Q-319 & S20240529S0149 &6408854408437947008& MSS\_47008 & 1 x 1700 & -16.18 ± 0.02 & \multicolumn{1}{c}{10.7} & 17.17\\
GS-2024A-Q-319 & S20240529S0150 &6491278576058838400& MSS\_38400 & 1 x 1100 & 241.28 ± 0.29 & \multicolumn{1}{c}{30.4} & 16.02\\
GS-2024A-Q-319 & S20240519S0156 &6810913576984559360& MSS\_59360 & 1 x 1100 & -321.34 ± 0.18 & \multicolumn{1}{c}{23.2} & 15.96\\
GS-2024A-Q-319 & S20240519S0155 & 6902720706255633664 & MSS\_33664 & 1 x 1300 & -286.78 ± 0.18 & \multicolumn{1}{c}{23.2} & 16.86\\
GS-2025B-Q-310 & S20250930S0065 &4690941913230909184& MSS\_09184 & 1 x 1300 & 156.98 ± 0.12 & \multicolumn{1}{c}{26.8} & 16.83\\
GS-2025B-Q-310 & S20251016S0055 S20251016S0056 S20251016S0057 S20251016S0058 S20251016S0059 &6578639928846550784& MSS\_50784 & 1 x 1800 1 x 1800 1 x 1800 1 x 1800 1 x 1800 & 29.99 ± 0.11 30.05 ± 0.03 30.04 ± 0.05 30.04 ± 0.14 30.05 ± 0.10 & \multicolumn{1}{c}{57.3} & 15.73\\
GS-2023B-FT-104 & S20231208S0058 &6500075906032035200& MSS\_35200 & 1 x 600 & 219.77 ± 0.48 & \multicolumn{1}{c}{26.0} & 16.21\\
GS-2024A-Q-319 & S20240530S0095 &6545400279293019008& MSS\_19008 & 1 x 1700 & -155.42 ± 0.14 & \multicolumn{1}{c}{16.5} & 17.17\\
\newpage
\multicolumn{1}{c}{\textbf{MW halo/wake}}\\
GS-2025B-Q-310 & S20251022S0017 S20251022S0018 S20251022S0019 S20251022S0020 S20251015S0039 &6903270668228195840& MSS\_95840 & 1 x 1800 1 x 1800 1 x 1800 1 x 1800 1 x 1800 & -266.95 ± 0.07 -266.99 ± 0.10 -266.97 ± 0.09 -266.95 ± 0.09 -267.05 ± 0.16 & \multicolumn{1}{c}{54.9} & 15.63\\
GS-2025B-Q-310 & S20251022S0021 S20251022S0022 S20251022S0023 S20251022S0024 S20251022S0025 S20251022S0026 &4923149526247840640 & MSS\_40640 & 1 x 1800 1 x 1800 1 x 1800 1 x 1800 1 x 1800 1 x 1800 & 226.01 ± 0.08 226.02 ± 0.05 226.03 ± 0.07 226.02 ± 0.10 226.06 ± 0.09 226.05 ± 0.06 & \multicolumn{1}{c}{95.0} & 15.89\\
GS-2025B-Q-310 & S20250930S0063 &4706631669280837248& MSS\_37248 & 1 x 900 & 86.86 ± 0.10 & \multicolumn{1}{c}{27.1} & 16.54\\
GS-2025B-Q-310 & S20251026S0043 &4905669009352040448& MSS\_40448 & 1 x 1300 & -91.42 ± 0.10 & \multicolumn{1}{c}{31.6} & 16.95\\
GS-2024A-Q-319 & S20240530S0094 &6908181396394873600& MSS\_73600 & 1 x 1500 & -375.97 ± 0.12 & \multicolumn{1}{c}{29.6} & 17.04\\
GS-2024A-Q-221 GS-2025B-Q-310 & S20240627S0114 S20240627S0115 S20240627S0117 S20251015S0040 S20250926S0034 &6788174100271623424& MSS\_23424 & 1 x 1750 1 x 1750 1 x 1750 1 x 971.0       1 x 971.0 & -37.31 ± 0.21 -37.21 ± 0.15 -37.27 ± 0.07 -37.93 ± 0.25 -38.31 ± 0.41 & \multicolumn{1}{c}{53.0} & 16.65\\
GS-2025B-Q-310 & S20250813S0119 &6498000990151085696& MSS\_85696 & 1 x 900 & 67.22 ± 0.11 & \multicolumn{1}{c}{46.1} & 16.02\\
GS-2024A-Q-221 GS-2023B-FT-104 & S20240629S0137 S20240629S0138 S20240629S0139 S20240629S0140 S20231209S0001 &6507990809203290368& MSS\_90368 & 1 x 1750 1 x 1750 1 x 1750 1 x 1750 1 x 800 & 224.86 ± 0.11 224.95 ± 0.09 224.87 ± 0.07 225.00 ± 0.15 224.94 ± 0.18 & \multicolumn{1}{c}{68.3} & 16.64\\
GS-2023B-FT-104 & S20231208S0059 &6554181430124375936& MSS\_75936 & 1 x 800 & -235.03 ± 0.09 & \multicolumn{1}{c}{24.7} & 16.4\\
GS-2024A-Q-319 & S20240529S0147 &6452441557622974336& MSS\_74336 & 1 x 1100 & -25.35 ± 0.13 & \multicolumn{1}{c}{22.3} & 16.71\\
GS-2024A-Q-319 & S20240529S0148 &6460034510046368000& MSS\_68000 & 1 x 1100 & 138.30 ± 0.14 & \multicolumn{1}{c}{9.6} & 16.8\\
GS-2024A-Q-319 & S20240529S0151 &6398469211875052544& MSS\_52544 & 1 x 1300 & -41.63 ± 0.23 & \multicolumn{1}{c}{16.6} & 16.94\\
GS-2025B-Q-310 & S20250913S0056 &2692140476715314304& MSS\_14304 & 1 x 900 & -19.75 ± 0.23 & \multicolumn{1}{c}{28.9} & 16.29\\
\multicolumn{1}{c}{\textbf{OC stream}}\\
GS-2025B-Q-310 & S20251118S0014 S20251119S0036 S20251119S0037 S20251119S0038 S20251119S0039 &6462341629039072896& MSS\_72896 & 1 x 1800 1 x 1800 1 x 1800 1 x 1800 1 x 1800 & -90.31 ± 0.10 -90.43 ± 0.14 -90.41 ± 0.18 -90.38 ± 0.11 -90.34 ± 0.06 & \multicolumn{1}{c}{76.3} & 15.94\\
GS-2025B-Q-310 & S20251016S0069 &6505337859444498688& MSS\_98688 & 1 x 900 & -83.47 ± 0.13 & \multicolumn{1}{c}{24.3} & 16.56\\
GS-2025B-Q-310 & S20250912S0021 &6779255411703094656& MSS\_94656 & 1 x 1100 & -133.42 ± 0.18 & \multicolumn{1}{c}{31.1} & 16.77\\
GS-2024A-Q-319 & S20240530S0093 &6785860826587863040& MSS\_63040 & 1 x 1100 & -87.89 ± 0.22 & \multicolumn{1}{c}{20.2} & 16.68\\
\multicolumn{1}{c}{\textbf{Nearby giants}}\\
GS-2025B-Q-310 & S20250913S0057 &6880996005762433664& MSS\_433664 & 1 x 900 & -9.75 ± 0.15 & \multicolumn{1}{c}{22.3} & 16.65\\
GS-2025B-Q-310 & S20250813S0098 &6406319587459014912& MSS\_14912 & 1 x 971 & 216.49 ± 0.09 & \multicolumn{1}{c}{20.3} & 16.79\\
GS-2025B-Q-310 & S20250912S0020 &6899708353634042752& MSS\_42752 & 1 x 1300 & -208.93 ± 0.12 & \multicolumn{1}{c}{35.1} & 17.09\\
GS-2025B-Q-310 & S20251016S0070         S20251016S0071 &4692493079322662528 & MSS\_62528 & 1 x 1000 1 x 1000 & 218.07 ± 0.18 218.19 ± 0.09 & \multicolumn{1}{c}{25.4} & 17.22\\
\hline
\hline
\end{longtable}\label{appendixtable2}
}

\end{appendix}

\end{document}